\pdfoutput=1
\documentclass[acmsmall,screen]{acmart}
\usepackage[utf8]{inputenc}
\usepackage{booktabs}
\usepackage{subcaption}
\usepackage{listings}
\usepackage{languages}
\usepackage{graphicx,wrapfig}
\usepackage{xspace}
\usepackage{color}
\usepackage{todonotes}
\usepackage{multirow}
\usepackage{paralist}
\usepackage{hyperref}
\usepackage{enumitem}
\usepackage{tabularx}

\newcommand{\Ie}{\emph{i.e.,}\xspace}
\newcommand{\Eg}{\emph{e.g.,}\xspace}

\newcommand{\Cf}{\emph{cf.}\xspace}
\renewcommand{\c}[1]{\lstinline|#1|\xspace}
\newcommand{\GH}{GitHub\xspace}
\newcommand{\Scala}{Scala\xspace}
\newcommand{\Java}{\textsc{Java}\xspace}
\newcommand{\SDB}{\textsc{SemanticDB}\xspace}
\newcommand{\SBT}{\textsc{Sbt}\xspace}
\newcommand{\ScalaMeta}{\textsc{Scalameta}\xspace}

\newcommand{\Scaladex}{Scaladex\xspace}

\newcommand{\Circe}{\c{circe.io}}
\newcommand{\Scalac}{\c{scalac}}
\newcommand{\ScalaJS}{\c{scalajs}}
\newcommand{\ScalaJSReact}{\c{scalajs-react}}
\newcommand{\Finagle}{\c{finagle}}
\newcommand{\Dotty}{\c{dotty}}
\newcommand{\Akka}{\c{akka}}
\newcommand{\Spark}{\c{spark}}
\newcommand{\Slinky}{\c{slinky}}
\newcommand{\ScalaTest}{\c{scalatest}}
\newcommand{\Specs}{\c{specs2}}
\newcommand{\Scalaz}{\c{scalaz}}

\newcommand{\TgDownloadedProjects}{65,177\xspace}
\newcommand{\TgDownloadedProjectsCodeRnd}{121.4M\xspace}

\newcommand{\TgSbtProjectsRnd}{43K\xspace}

\newcommand{\TgMavenProjectsRnd}{5.1K\xspace}

\newcommand{\TgGraddleProjectsRnd}{1.5K\xspace}

\newcommand{\TgCompatibleSbtProjectsRnd}{23.6K\xspace}

\newcommand{\TgDuplicated}{12,550\xspace}
\newcommand{\TgDuplicatedCodeRnd}{33.1M\xspace}

\newcommand{\TgSelectedProjects}{11,057\xspace}

\newcommand{\TgExtractedProjects}{7,326\xspace}

\newcommand{\TgLostStarsInDuplicatedProjectsPct}{2.8\%\xspace}

\newcommand{\TgCorpus}{7,280\xspace}
\newcommand{\TgCorpusCodeRnd}{18.7M\xspace}

\newcommand{\TgCorpusGencodeRnd}{2.2M\xspace}

\newcommand{\TgCorpusTestCodeRnd}{5.9M\xspace}

\newcommand{\TgCorpusCodeMedian}{677\xspace}

\newcommand{\TgCorpusScalaEleven}{4,197\xspace}

\newcommand{\TgCorpusScalaElevenCodePct}{43.8\%\xspace}
\newcommand{\TgCorpusScalaElevenStarsPct}{33.7\%\xspace}

\newcommand{\TgSparkProjectsBefore}{102\xspace}

\newcommand{\TgSparkProjectsBeforeCodePct}{37.6\%\xspace}

\newcommand{\TgFailedProjectsAll}{3,731\xspace}

\newcommand{\TgFailedProjectsMissingDependenciesRnd}{2.1K\xspace}

\newcommand{\TgFailedProjectsCompileError}{873\xspace}
\newcommand{\TgFailedProjectsBrokenBuild}{189\xspace}
\newcommand{\TgFailedProjectsEmptyBuild}{156\xspace}
\newcommand{\TgFailedProjectsMissingScalajs}{964\xspace}
\newcommand{\TgFailedProjectsMissingSnapshots}{263\xspace}

\newcommand{\TgDataCleanupRemovedProjects}{46\xspace}

\newcommand{\TgDataCleanupProjectsLostCodePct}{1.1\%\xspace}

\newcommand{\TgImplicitDeclarationsRnd}{370.7K\xspace}

\newcommand{\TgImplicitDeclarationsPublicRnd}{332.2K\xspace}

\newcommand{\TgImplicitDeclarationsPublicPct}{89.6\%\xspace}

\newcommand{\TgCallSitesRnd}{29.6M\xspace}

\newcommand{\TgImplicitCallSitesRnd}{8.1M\xspace}

\newcommand{\TgImplicitCallSitesPct}{27.2\%\xspace}

\newcommand{\TgProjectsUsingImplicits}{7,148\xspace}
\newcommand{\TgProjectsDefiningImplicitsRnd}{5.7K\xspace}

\newcommand{\TgProjectsUsingImplicitsPct}{98.2\%\xspace}
\newcommand{\TgProjectsDefiningImplicitsPct}{78.2\%\xspace}
\newcommand{\TgProjectsUsingImplicitsRndPct}{98\%\xspace}
\newcommand{\TgProjectsDefiningImplicitsRndPct}{78\%\xspace}
\newcommand{\TgImplicitCallSitesRndPct}{27\%\xspace}

\newcommand{\TgProjectsTargetingJvm}{7,075\xspace}

\newcommand{\TgCtxPrettifierRnd}{563.6K\xspace}

\newcommand{\TgCtxPretrifierProjectsRnd}{2.5K\xspace}

\newcommand{\TgIcAllProjectsUsing}{7,050\xspace}
\newcommand{\TgIcAllProjectsUsingPct}{96.8\%\xspace}

\newcommand{\TgIcAllProjectsDeclaring}{2,991\xspace}
\newcommand{\TgIcAllProjectsDeclaringPct}{41.1\%\xspace}
\newcommand{\TgIcAllCallsitesRnd}{2.5M\xspace}

\newcommand{\TgIcAllCallsitesPct}{31.5\%\xspace}

\newcommand{\TgIcAllDeclarations}{61,995\xspace}
\newcommand{\TgIcAllDeclarationsPct}{16.7\%\xspace}

\newcommand{\TgIcTestCallsitesPct}{61.1\%\xspace}

\newcommand{\TgIcCallsitesFromScalaInMainPct}{47.4\%\xspace}
\newcommand{\TgIcCallsitesFromTestLibrariesPct}{59.4\%\xspace}
\newcommand{\TgIcCallsitesFromLocalDeclarationsPct}{18.8\%\xspace}

\newcommand{\TgIcDeclarationsSd}{615.5\xspace}
\newcommand{\TgIcDeclarationsMedian}{3\xspace}

\newcommand{\TgIcUsedInJsVsJvm}{2.5\xspace}

\newcommand{\TgIcInScalatest}{343\xspace}
\newcommand{\TgIcInSpec}{440\xspace}

\newcommand{\TgIcInSlinkyRnd}{33.6K\xspace}

\newcommand{\TgIcSlinkyUse}{8\xspace}
\newcommand{\TgIcSlinkyUseProjects}{2\xspace}
\newcommand{\TgIcSlinkyUseProjectsCodeRnd}{3.6K\xspace}

\newcommand{\TgIcfunAllDeclarationsRnd}{1.1K\xspace}

\newcommand{\TgIcfunAllDeclarationsPct}{0.3\%\xspace}

\newcommand{\TgImplicitsCallsitesPctMedian}{23.4\%\xspace}
\newcommand{\TgImplicitsTestCallsitesPctMedian}{38\%\xspace}
\newcommand{\TgImplicitsMainCallsitesPctMedian}{17.1\%\xspace}
\newcommand{\TgIpInCallSitesPct}{60.3\%\xspace}

\newcommand{\TgIpScalaCollectionAppBigProjectsUsingPct}{96.6\%\xspace}

\newcommand{\TgIpScalaCollectionTestProjectsUsingPct}{44.6\%\xspace}
\newcommand{\TgIpScalaCollectionAllProjectsUsingPct}{80.3\%\xspace}

\newcommand{\TgIpScalaCollectionLibCallsitesPct}{38.3\%\xspace}

\newcommand{\TgIpScalaReflectionAppBigProjectsUsingPct}{90.2\%\xspace}
\newcommand{\TgIpScalaReflectionAppSmallProjectsUsingPct}{56\%\xspace}
\newcommand{\TgIpScalaReflectionLibProjectsUsingPct}{61.7\%\xspace}
\newcommand{\TgIpScalaReflectionTestProjectsUsingPct}{56.4\%\xspace}

\newcommand{\TgIpScalaConcurrencyAppBigProjectsUsingPct}{58.5\%\xspace}
\newcommand{\TgIpScalaConcurrencyAppSmallProjectsUsingPct}{32.1\%\xspace}

\newcommand{\TgCallsitesFormSlick}{2.6\%\xspace}
\newcommand{\TgCallsitesFormPlay}{5\%\xspace}
\newcommand{\TgCallsitesFormAkka}{2.3\%\xspace}
\newcommand{\TgUseScalacticSourcePositionRnd}{912.7K\xspace}

\newcommand{\TgIpCallsitesRnd}{3.7M\xspace}

\newcommand{\TgIpCallsitesPct}{46.2\%\xspace}

\newcommand{\TgLtConditionalPct}{16.4\%\xspace}

\newcommand{\TgLtGoodPct}{79.8\%\xspace}

\newcommand{\TgLtJavaPct}{19.7\%\xspace}
\newcommand{\TgLtJavaIo}{53\xspace}
\newcommand{\TgLtJavaDate}{27\xspace}
\newcommand{\TgLtJavaLang}{50\xspace}

\newcommand{\TgLtJavaLibraries}{176\xspace}
\newcommand{\TgLtFromPrimitive}{990\xspace}

\newcommand{\TgLtFromString}{217\xspace}
\newcommand{\TgLtFromInt}{77\xspace}

\newcommand{\TgEmClassOverDef}{12,150\xspace}
\newcommand{\TgEmClassOverDefPct}{65.3\%\xspace}
\newcommand{\TgEmConditionalRnd}{1.9K\xspace}

\newcommand{\TgEmConditionalPct}{10.2\%\xspace}

\newcommand{\TgEmJavaRnd}{1.7K\xspace}

\newcommand{\TgEmJavaPct}{9.3\%\xspace}
\newcommand{\TgEmJavaIo}{224\xspace}
\newcommand{\TgEmJavaDate}{200\xspace}
\newcommand{\TgEmJavaLang}{59\xspace}

\newcommand{\TgEmJavaLibraries}{676\xspace}
\newcommand{\TgEmFromPrimitiveRnd}{3.7K\xspace}

\newcommand{\TgEmFromString}{1,169\xspace}
\newcommand{\TgEmFromInt}{452\xspace}
\newcommand{\TgEmConditionalTcRnd}{1.6K\xspace}

\newcommand{\TgEmConditionalCtx}{323\xspace}
\newcommand{\TgCtxJdkRnd}{1.8K\xspace}

\newcommand{\TgCtxJdkTypes}{50\xspace}
\newcommand{\TgCtxJdkProjects}{179\xspace}

\newcommand{\TgCtxPrimitive}{1,044\xspace}
\newcommand{\TgCtxPrimitiveProjects}{159\xspace}
\newcommand{\TgCtxFunction}{645\xspace}
\newcommand{\TgCtxFunctionProjects}{154\xspace}
\newcommand{\TgUicDefinedRnd}{41.5K\xspace}

\newcommand{\TgUicDefinedInSlinkyRnd}{33.6K\xspace}

\newcommand{\TgUicDefiningProjectsRnd}{1.2K\xspace}

\newcommand{\TgUicDefiningProjectsPct}{16.2\%\xspace}
\newcommand{\TgUicUsingProjectsRnd}{6.1K\xspace}

\newcommand{\TgUicUsingProjectsPct}{83.9\%\xspace}
\newcommand{\TgUicDefinedInLibsRnd}{1.9K\xspace}

\newcommand{\TgUicDefinedInLibsProjects}{619\xspace}
\newcommand{\TgUicPrimitiveRnd}{1.6K\xspace}

\newcommand{\TgUicPrimitiveProjects}{552\xspace}

\newcommand{\TgUicBothPrimitive}{81\xspace}
\newcommand{\TgUicBothPrimitiveProjects}{47\xspace}
\newcommand{\TgBicDefinedRnd}{1.1K\xspace}

\newcommand{\TgBicDefiningProjects}{209\xspace}
\newcommand{\TgBicDefiningProjectsPct}{2.9\%\xspace}
\newcommand{\TgBicUsingProjectsRnd}{1.9K\xspace}

\newcommand{\TgBicUsingProjectsPct}{26.5\%\xspace}
\newcommand{\TgBicFromOrToPrimitives}{129\xspace}
\newcommand{\TgBicFromAndToPrimitives}{13\xspace}

\newcommand{\TgBicProjectsUsingDecorateAsPct}{22.4\%\xspace}
\newcommand{\TgBicProjectsUsingWrapAs}{728\xspace}
\newcommand{\TgBicProjectsUsingWrapAsPct}{10\%\xspace}
\newcommand{\TgBicProjectsUsingBoth}{244\xspace}
\newcommand{\TgImplicitNotFoundUseRnd}{110.9K\xspace}

\newcommand{\TgImplicitNotFoundTypesRnd}{1.2K\xspace}

\newcommand{\TgImplicitNotFoundProjects}{436\xspace}

\newcommand{\TgLengthOfInjectedDeclarationsRnd}{55M\xspace}
\newcommand{\TgLengthOfInjectedDeclarations}{3.5\xspace}

\newcommand{\TgCsNestingMedian}{1\xspace}

\newcommand{\TgCsNestingMax}{5,695\xspace}
\newcommand{\TgCsNestingMaxProject}{xdotai/typeless\xspace}
\newcommand{\TgCsNestingMaxCharLengthRnd}{56.2K\xspace}

\newcommand{\CompilationTimeProjects}{1,969\xspace}

\newcommand{\CompilationTimeProjectsCode}{8.4M\xspace}

\newcommand{\CompilationTimeProjectsTC}{488\xspace}

\newcommand{\CompilationTimeProjectsTCCode}{2.8M\xspace}

\setcopyright{rightsretained}
\acmPrice{}
\acmDOI{10.1145/3360589}
\acmYear{2019}
\copyrightyear{2019}
\acmJournal{PACMPL}
\acmVolume{3}
\acmNumber{OOPSLA}
\acmArticle{163}
\acmMonth{10}

\begin{document}
\title{Scala Implicits Are Everywhere}
\subtitle{A Large-Scale Study of the Use of Scala Implicits in the Wild}


\author{Filip Křikava}\affiliation{\institution{Czech Technical University in Prague}\country{CZ}}
\author{Heather Miller}\affiliation{\institution{Carnegie Mellon University}\country{USA}}
\author{Jan Vitek}\affiliation{\institution{Czech Technical University in Prague and Northeastern University}\country{USA}}
\authorsaddresses{}
\renewcommand{\shortauthors}{Krikava, Miller, Vitek}

\begin{abstract}
  The \Scala programming language offers two distinctive language features
  \emph{implicit parameters} and \emph{implicit conversions}, often referred
  together as \emph{implicits}. Announced without fanfare in 2004, implicits
  have quickly grown to become a widely and pervasively used feature of the
  language. They provide a way to reduce the boilerplate code in Scala
  programs. They are also used to implement certain language features without
  having to modify the compiler. We report on a large-scale study of the use of
  implicits in the wild. For this, we analyzed \TgCorpus \Scala
  projects hosted on \GH, spanning over \TgImplicitCallSitesRnd call sites
  involving implicits and \TgImplicitDeclarationsRnd implicit declarations
  across \TgCorpusCodeRnd lines of \Scala code.
\end{abstract}
\maketitle

\section{Introduction}

\begin{tabular}{lrr}
\footnotesize\it ``...experienced users claim that code
    bases are train wrecks because of overzealous use of
    implicits.'' & ~~~&\tiny  --M. Odersky, 2017\\[0mm]
\footnotesize \it ``...can impair readability or introduce
    surprising behavior, because  of a subtle chain of
    inference.''& &\tiny --A. Turon, 2017\\[0mm]
\footnotesize \it ``Any sufficiently advanced technology is
    indistinguishable from magic.''& &\tiny --A.C. Clarke, 1962\\[2mm]
\end{tabular}

\noindent
Programming language designers strive to find ways for their users to
express programming tasks in ways that are both concise and readable. One
approach to reduce boilerplate code is to lean on the compiler and its
knowledge and understanding of the program to fill in the ``boring parts''
of the code.  The idea of having the compiler automatically provide missing
arguments to a function call was first explored by~\citet{lewis00} in
Haskell and later popularized by Scala as {\em implicit parameters}. {\em Implicit
conversions} are related, as they rely on the compiler to automatically adapt
data structures in order to avoid cumbersome explicit calls to constructors.
For example, consider the following code snippet:
~ ~
\c{
    "Just like magic!".enEspanol} ~ ~
Without additional context one would expect the code not to compile as the
\c{String} class does not have a method \c{enEspanol}. In \Scala, if the
compiler is able to find a method to convert a string object to an instance
of a class that has the required method (which resolves the type error), that
conversion will be inserted silently by the compiler and, at runtime, the method
will be invoked to return a value, perhaps \c{"Como por arte de magia!"}.

Implicit parameters and conversions provide ways to (1) extend existing
software~\cite{Lammel2006-sk} and implement language features outside of the
compiler~\cite{Miller2013-nq}, and (2) allow end-users to write code with less
boilerplate~\cite{haoyi16_patterns}. They offload the task of selecting and
passing arguments to functions and converting between types to the compiler. For
example, the \c{enEspanol} method from above uses an implicit parameter to get a
reference to a service that can do the translation:
\hspace{.1cm} \c{ def enEspanol(implicit ts:Translator):String}.
Calling a function that has implicit arguments results in the omitted
arguments being filled from the context of the call based on their
types. Similarly, with an implicit conversion in scope, one can seamlessly
pass around types that would have to be otherwise converted by the
programmer.

\begin{figure}[t!]\begin{tabular}{@{}cc@{}}
\begin{minipage}{.65\linewidth}
  \includegraphics[width=\linewidth]{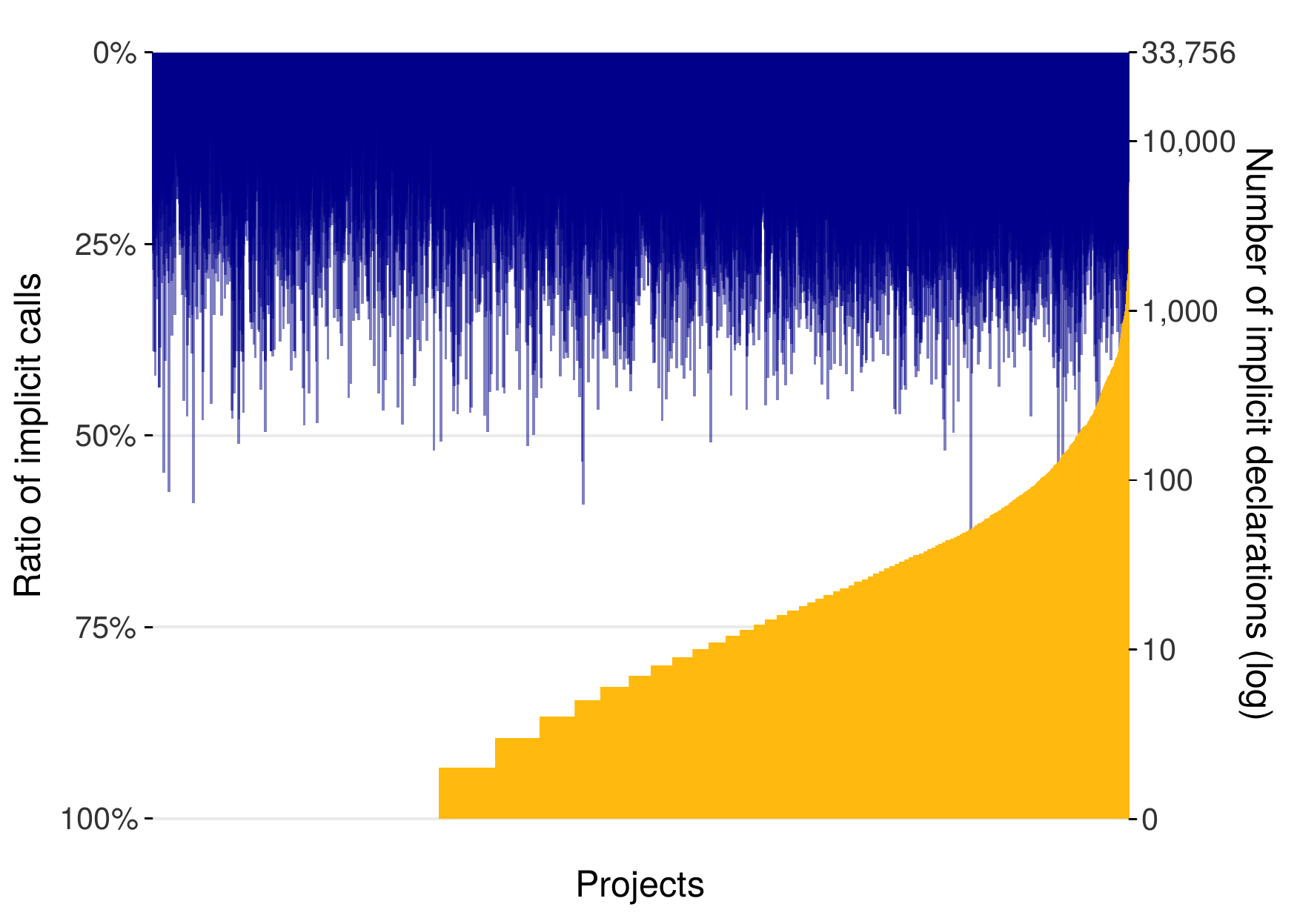}
\end{minipage}&\begin{minipage}{.32\linewidth} \small\vspace{-2mm}
  {\underline{\small At a glance:}}\\
  \begin{compactitem}[$-$]
  	\item \TgCorpus \Scala projects
  	\item \TgCorpusCodeRnd lines of code
  	\item \TgImplicitCallSitesRnd implicit call sites
  	\item \TgImplicitDeclarationsRnd implicit declarations
  \end{compactitem}
  \begin{flushleft}
\small
 	\vspace{1mm} {\bf \TgProjectsUsingImplicitsRndPct of projects use
          implicits}\\ {\bf \TgProjectsDefiningImplicitsRndPct of projects
          define implicits}\\ {\bf \TgImplicitCallSitesRndPct of call sites
          use implicits}
    \vspace{1mm}
  \end{flushleft}
The top of the graph shows the ratio of call sites, in each project, that
involves implicit resolution.  The bottom shows the number of implicit
definitions in each project.
\end{minipage} \end{tabular}
\caption{Implicits usage across our corpus} \label{fig:implicits-overview}
\vspace*{-5mm}
\end{figure}

\paragraph{The Good: A Powerful Tool.} It is uncontroversial to assert
that implicits changed how \Scala is used. Implicits gave rise to new coding
idioms and patterns, such as type classes~\cite{oliveira10_type_classes}. They
are one of a few key features which enable embedding Domain-Specific Languages
(DSLs) in Scala. They can be used to establish or pass context (e.g., implicit
reuse of the same threadpool in some scope), or for dependency injection.
Implicits have even been used for
computing new types and proving relationships between them~\cite{shapeless,
Miller2014-nu}.
The \Scala community adopted implicits enthusiastically and uses them to solve a
host of problems. Some solutions gained popularity and become part of the
unofficial programming lexicon. As usage grew, the community endeavored to
document and teach these idioms and patterns by means of blog
posts~\cite{haoyi16_patterns}, talks~\cite{odersky17_scaladays} and the official
documentation~\cite{suereth13_implicit_classes}. While these idioms are believed
to be in widespread use, there is no hard data on their adoption. How widespread
is this language feature? And what do people do with
implicits? Much of our knowledge is folklore based on
a handful of popular libraries and discussion on various shared forums.

Our goal is to document, for language designers and software
engineers, how this feature is really used in the wild, using a large-scale
corpus of real-world programs.  We provide data on how they are used in popular
projects engineered by expert programmers as well as in projects that are likely
more representative of how the majority of developers use the language. This
paper is both a retrospective on the result of introducing this
feature into the wild, as well as a means to inform designers of future
language of how people use and misuse implicits.

\paragraph{The Bad: Performance} While powerful, implicits aren't without
flaws. Implicits have been observed to affect
compile-time performance; sometimes significantly.
For example, a popular \Scala project reported a three order-of-magnitude
speed-up when developers realized that an implicit conversion was silently
converting \Scala collections to Java collections only to perform a single
operation that should have been done on the original object.\footnote{Documented
in
\url{https://github.com/mesosphere/marathon/commit/fbf7f29468bda2ec29b7fbf80b6864f46a825b7a}.}
Another project reported a 56 line file taking 5 seconds to compile because
of implicit resolution. Changing one line of code to remove an implicits,
improved compile time to a tenth of second~\cite{torreborre17}.  Meanwhile,
faster compilation is the most wished-for improvement for future releases of
Scala~\cite{scala-dev-survey2018}. Could implicit resolution be a
significant factor affecting compilation times across the Scala ecosystem?

\begin{figure}[t!]\begin{tabular}{p{0.5\textwidth}|p{0.47\textwidth}}
\begin{minipage}{.5\textwidth}\centering\begin{lstlisting}
  case class Card(n:Int, suit:String) {
    def isInDeck(implicit deck: List[Card]) =
      deck contains this
  }
  implicit def intToCard(n:Int) = Card(n, "club")
  implicit val deck = List(Card(1, "club"))

  1.isInDeck
\end{lstlisting}\end{minipage}&
\begin{minipage}{.47\textwidth}\centering
\includegraphics[width=.9\linewidth]{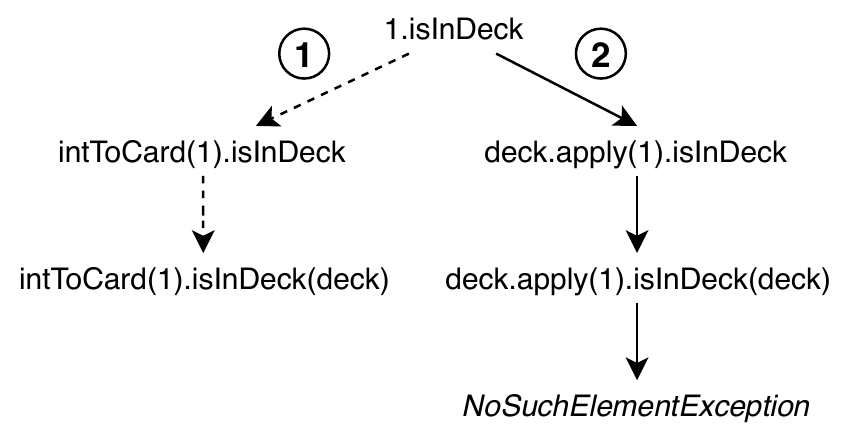}
\end{minipage}\end{tabular}
\caption{Instead of injecting a conversion to \c{intToCard} (1), the
  compiler injects \c{deck.apply} (2) since \c{List[A]} extends
  (transitively) \c{Function[Int,A]}.  An exception is thrown because the
  \c{deck} contains only one element  (\url{http://scalapuzzlers.com/})} \label{fig:confusing-implicit-derivation}
\end{figure}

\paragraph{The Ugly: Readability} Anecdotally, there are signs that the
design of implicits can lead to confusing scenarios or
difficult-to-understand code. Figure~\ref{fig:confusing-implicit-derivation}
illustrates how understanding implicit-heavy code can place an unreasonable
burden on programmers\footnote{For example, an entire book is devoted to
  so-called ``puzzlers,'' or ``enigmatic Scala code that behave highly
  contrary to expectations'' which ``will entertain and enlighten even the
  most accomplished developer''~\cite{ScalaPuzzlers}}. In this
example, the derivation chosen by the compiler leads to an error which
requires understanding multiple levels of the type hierarchy of the \c{List}
class.
Such readability issues have even lead the Scala creators to reconsider the
design of Scala's API-generation tool, Scaladoc. This was due to community
backlash~\cite{longestsuicidenote} following the introduction of the Scala 2.8
Collections library~\cite{Odersky2009-zi}---a design which made heavy use of
implicits in an effort to reduce code duplication. The design caused a
proliferation of complex method signatures across common data types throughout
the Scala standard library, such as the following implementation of the \c{map}
method which was displayed by Scaladoc as:
~\,~\c{def map[B,That](f:A=>B)(implicit bf:CanBuildFrom[Repr,B,That]):That.} ~\,~
To remedy this, Scaladoc was updated with {\em
use-cases},\footnote{\Cf \url{https://docs.scala-lang.org/overviews/scaladoc/for-library-authors.html}}
a feature designed to allow library authors to manually override method
signatures with simpler ones in the interest of hiding complex type signatures
often further complicated by implicits. The same \c{map} signature
thus appears as follows in Scaladoc after simplification with a \c{@usecase}
annotation: \hspace{.2cm} \c{    def map[B](f: (A) => B): List[B]}

\paragraph{\bf This Work} To understand the use of implicits across the
\Scala ecosystem, we have built an open source and reusable pipeline to
automate the analysis of large \Scala code bases, compute statistics and
visualize results.  We acquired and processed a corpus of \TgCorpus projects
from \GH with over \TgImplicitCallSitesRnd implicit call sites and more than
\TgImplicitDeclarationsRnd implicit declarations across \TgCorpusCodeRnd
non-empty lines of \Scala code. We observed over
\TgProjectsUsingImplicitsPct projects using implicits, and
\TgProjectsDefiningImplicitsPct projects declaring implicits. With close to
\TgImplicitCallSitesPct of call sites requiring implicit resolution,
implicits are the most used feature of \Scala.
Figure~\ref{fig:implicits-overview} summarizes the usage of implicits in our
corpus.  Our results document which idioms and patterns are popular and in
application, library and tests. We provide data on the compilation time cost
of implicits and the complexity of implicits. Our artifact is available at:

\begin{center}
\url{https://doi.org/10.5281/zenodo.3369436}
\end{center}

\section{An Overview of Scala Implicits}\label{sec:scala-implicits}

\Scala is a statically typed language that bridges the gap between
object-oriented and functional programming. Implicits were included
in the first release in 2004. In that version
\emph{implicit conversions} were used to solve the late extension problem;
namely, given a class $C$ and a trait $T$, how to have $C$ extend $T$
without touching or recompiling $C$. Conversions add a wrapper when a member of $T$ is
requested from an instance of $C$. \Scala 2.0 added \emph{implicit
  parameters} in 2006.

\subsection{Implicit Conversions}\label{sec:implicit-conversion}

Implicit conversion provides a way to use a type where another type is
required without resorting to an explicit conversion. They are applied when
an expression does not conform to the type expected by its context or when a
called method is not defined on the receiver type. A conversion is defined
with an implicit function or class, or an implicit value of a function type
(\Eg \c{implicit val x:A=>B}).

Implicit conversions are not specific to \Scala. They also appear in
languages such as C++ or C\#.  The difference is that conversions are
typically defined in the class participating in the conversion, while in
\Scala the implicit conversions can be defined in types unrelated to the
conversion types. This allows programmers to selectively import
conversions. For example it is possible to define an implicit conversion
from a \c{String} to an \c{Int}:
\begin{lstlisting}
  implicit def string2int(a: String): Int = Integer.parseInt(a)
  val x: Int = "2"
\end{lstlisting}
\noindent
Implicit conversions are essential to provide seamless interoperability with
Java which was important in the early days of \Scala. Conversions are also one
of the main building blocks for constructing embedded Domain-Specific
Languages (DSLs). For example, the following code snippet adds some simple time
unit arithmetic that feels natural in the language.
\begin{lstlisting}
    case class Duration(time: Long, unit: TimeUnit) {
      def +(o: Duration) = Duration(time + unit.convert(o.time, o.unit), unit)
    }
    implicit class Int2Duration(that: Int) {
      def seconds = new Duration(that, SECONDS); def minutes = new Duration(that, MINUTES)
    }
    5.seconds + 2.minutes //Duration(125L, SECONDS)
\end{lstlisting}

\subsection{Implicit Parameters}

A method or a constructor can define implicit parameters. The arguments to
these parameters will be filled in by the compiler at every call site with the
most suitable values in the calling context. For example, a function \c{def
  sub(x:Int)(implicit y:Int)=x-y} with implicit parameter \c{y} can be called
with \c{sub(1)} provided that the compiler can find an implicit such as
\c{implicit val n=1}. The compiler looks for implicits in the current lexical
scope and if there are no eligible identifiers then it searches the implicit
scope of the implicit parameter's type (associated companion objects\footnote{A
  companion object is a singleton associated with a class used to define static
  fields and methods.} and packages). If a value is found, the compiler injects
it into the argument list of the call. If multiple values are found and none of
them is more specific than the others, an ambiguity compilation error is
thrown. An error is also raised when no eligible candidate is found.
Importantly, besides having the correct type, an implicit value from a lexical
scope must be accessible using its simple name (without selecting from another
value using dotted syntax). This means that normal rules for name binding
including shadowing apply. Implicit values (\c{val}), variables (\c{var}),
objects (\c{object}) and functions (\c{def}) without explicit parameters can
all be used to fill implicit parameters. An implicit parameter of a function
type \c{A => B} can be used as an implicit conversion in the method body. For
example \c{def get[T, C](xs: C, n: Int)(implicit conv: C => Seq[T]):T = xs(n)}
can be called with any type \c{C}, as long as there is implicit conversion in
scope that can convert \c{C} into a sequence.

\subsection{Idioms and Patterns}

Over time, programmers have put implicits to many uses. This section describes
the most widely discussed implicit idioms. This list is based on our
understanding of the state of practice. It is not expected to be exhaustive or
definitive.

\subsubsection{Late Trait Implementation}

This idiom is a solution for the late extension problem, and was the
original motivation for adding implicits to Scala in the first place. To add a new
trait to an existing class, one can define a one-parameter conversion that
returns an instance of the trait.
\begin{lstlisting}
  implicit def call2Run(x: Callable[_]): Runnable = new Runnable { def run = x.call }
\end{lstlisting}
This snippet adds the \c{Runnable} interface to any any type that implements
\c{Callable}. Conversions can also take implicit parameters, they are then
referred to as \emph{conditional} conversions.
\begin{lstlisting}
  implicit def call2Future[T](x: Callable[T])(implicit ctx: ExecutionContext): Future[T]
\end{lstlisting}
For example, the above defines a late trait implementation that is only applicable if there
exists an execution context in scope.

\subsubsection{Extension Methods}\label{sec:extension-methods}

Extension methods allow developers to add methods to existing classes. They
are defined with an \c{implicit def} that converts objects to a new class
that contains the desired methods. \Scala 2.10 added syntactic sugar to
combine conversion and class declaration in the \c{implicit class}
construct. The conversion takes a single non-implicit parameter as shown in
the following snippet where \c{zip} is added to any \c{Callable}.
\begin{lstlisting}[escapechar=]
  implicit class XtensionCallable[T](x: Callable[T]) {
    def zip[U](y: Callable[U]): Callable[(T, U)] = () => (x.call, y.call)
  }
  val c1 = () => 1; val c2 = () => true; val r = c1 zip c2 // r: Callable[(Int, Boolean)]
\end{lstlisting}

\noindent
An extension method is convenient as it allows to write \c{c1 zip c2}
instead of \c{zip(c1, c2)}. It is an important feature for embedded DSLs. On
the other hand, unlike static methods, it is harder to read.  Without
knowing the complete code base it is difficult to know where a calling
method is defined and how the definition got into the current scope.
Extension methods can also be conditional. For example, we can add a \c{def
  schedule(implicit c: ExecutionContext)} method that will run the callable on
the implicitly provided execution context if it is present. If there is none,
the developer will get a compile-time error ``\emph{cannot find an implicit
  ExecutionContext ... import scala.concurrent.ExecutionContext.global.}''.
This is because the \c{ExecutionContext} is annotated with
\c{@implicitNotFound}, a \Scala annotation allowing one to customize the
compile-time error message that should be outputted in the case no implicit
value of the annotated type is available.

\subsubsection{Type Classes}\label{sec:idioms-type-class}

\citet{oliveira10_type_classes} demonstrated how to use implicit parameters
to implement type
classes~\cite{wadler89_type_classes}. Fig.~\ref{fig:tc-example}a defines a
trait \c{Show} that abstracts over pretty-printing class instances.  The
function \c{show} can be called on instances \c{T}, for which there is an
implicit value of type \c{Show[T]}. This allows us to retrospectively add
support to classes we cannot modify. For example, given a class
\c{Shape(sides: Int)} from a 3rd party library, we can define the implicit
value \c{ShapeShow} to add pretty printing
(Fig.~\ref{fig:tc-example}b). This is an implicit object that extends
\c{Show} and implements \c{show}. Thus when \c{show} is called with an
explicit argument of type \c{Shape}, for example \c{show(Shape(5))}, the
compiler adds the implicit \c{ShapeShow} as the implicit argument \c{ev},
resulting in \c{show(Shape(5))(shapeShow)}.

\begin{figure}[h!]
\begin{minipage}[t]{.45\textwidth}
\begin{lstlisting}
  trait Show[T] {
    def show(x: T): String
  }


  def show[T](x: T)(implicit ev: Show[T]) = ev.show(x)
\end{lstlisting}
\centering \textsf{(a)}
\end{minipage}\hfill
\begin{minipage}[t]{.55\textwidth}
\begin{lstlisting}
  case class Shape(n: Int)

  implicit object shapeShow extends Show[Shape] {
    def show(x: Shape) = x.n match {
      case 3 => "a triangle"; case 4 => "a square"
      case _ => "a shape with $n sides" }
    }
\end{lstlisting}
\centering \textsf{(b)}
\end{minipage}
\begin{lstlisting}
  implicit def listShow[T](implicit ev: Show[T]) = new Show[List[T]] {
    def show(x: List[T]) = x.map(x => ev.show(x)).mkString("a list of [", ", ", "]")
  }
\end{lstlisting}
\centering \textsf{(c)}

\vspace{-3mm}\caption{Type classes}\label{fig:tc-example}\vspace{-1mm}
\end{figure}

\noindent
Since functions can be used as implicit parameters, we can generalize this
example and create an implicit allowing us to show a sequence of showable
instances. In the following snippet, \c{listShow} is a generic type class
instance that combined with an instance of type \c{Show[T]} returns a type
class instance of type \c{Show[List[T]]} (Fig.~\ref{fig:tc-example}c). Thus, a
call to \c{show(List(Shape(3), Shape(4)))} is transformed to
\c{show(List(Shape(3), Shape(4)))(listShow[Shape](shapeShow))}, with two levels
of implicits inserted. This implicit type class derivation is what makes type
classes very powerful. The mechanism can be further generalized using implicit
macros to define a \emph{default} implementation for type class instances that
do not provide their own specific ones~\cite{shapeless, Miller2014-nu}.

\subsubsection{Extension Syntax Methods}

Type classes define operations on types, when combined with extension
methods it is possible to bring these operations into the corresponding
model types. We can extend the \c{Show[T]} type class and define an
extension method
\begin{lstlisting}
  implicit class ShowOps[T](x: T)(implicit s: Show[T]) { def show = s.show(x) }
\end{lstlisting}
allowing one to write directly \c{Shape(3).show} instead of
\c{show(Shape(3))}.  The \c{ShowOps[T]} is a conditional conversion that is
only applied if there is an instance of the \c{Show[T]} in scope. This allows
library designers to use type class hierarchies instead of the regular
sub-typing. The name \emph{extension syntax methods} comes from the fact
that developers often lump these methods into a package called
\emph{syntax}.

\subsubsection{Type Proofs}

Implicit type parameters can used to enforce API rules at a compile time by
encoding them in types of implicit parameters. For example, \c{flatten} is a
method of \c{List[A]} such that given an instance \c{xs: List[List[B]]},
\c{xs.flatten} returns \c{List[B]} concatenating the nested lists into a
single one. This is done with an implicit parameter:
\begin{lstlisting}
  class List[A] { def flatten[B](implicit ev: A => List[B]): List[B] }
\end{lstlisting}
Here, \c{A => List[B]} is an implicit conversion from \c{A} to \c{List[B]}. It
can also be viewed as a predicate that must be satisfied at compile time in
order for this method to be called. We can define an implicit function
\c{implicit def isEq[A]: A=>A = new =>[A,A]\{\}} that will act as generator
of proofs such that \c{A} in \c{A => List[B]} is indeed \c{List[B]}. Therefore,
a call \c{List(List(1)).flatten} will be expanded to
\c{List(List(1)).flatten(isEq[List[Int]])} since \c{A} is a \c{List} while
\c{List(1).flatten} will throw a compile time exception: \emph{``No
  implicit view available from Int => List[B]''}.

\subsubsection{Contexts}

Implicit parameters can reduce the boilerplate of threading a context parameter
through a sequence of calls. For example, the methods in \c{scala.concurrent},
the concurrency library in Scala's standard library, all need an
\c{ExecutionContext} (e.g., a thread pool or event loop) to execute their tasks
upon. The following code shows the difference between explicit and implicit
contexts.
\begin{minipage}[b]{.49\textwidth}
\begin{lstlisting}[title=With \emph{explicit} context]
  val ctx = ExecutionContext.global
  val f1 = Future(1)(ctx)
  val f2 = Future(2)(ctx)
  val r = f1.flatMap(r1 =>
    f2.map(r2 => r1 + r2)(ctx)
  )(ctx)
\end{lstlisting}
\end{minipage}
\begin{minipage}[b]{.49\textwidth}
\begin{lstlisting}[title=With \emph{implicit} context]
  implicit val ctx = ExecutionContext.global
  val f1 = Future(1)
  val f2 = Future(2)


  val r = for(r1 <- f1; r2 <- f2) yield r1 + r2
\end{lstlisting}
\end{minipage}

\noindent On the left, an explicit context is passed around on every call to a
method on \c{Future}, while on the right much of the clutter is gone thanks to
implicits. This de-cluttering hides the parameters and makes calls to \c{map}
and \c{flatMap} more concise.  The idiom consists of the declaration of an
implicit context (usually as an \c{implicit val}), and the declaration of the
functions that handle it.

\subsubsection{Anti-patterns: Conversions}

A widely discussed anti-pattern is the conversions between types in unrelated
parts of the type hierarchy. The perceived danger is that any type can be
automatically coerced to a random type unexpectedly; \Eg imagine a conversion
from \c{Any} to \c{Int} introduced into the root of a big project. One could
imagine such a conversion wreaking havoc in surprising places in a code base and
being difficult to track down.  Another anti-pattern is conversions that go both
ways~\cite{odersky17_scaladays}. Since conversions are not visible, it is
difficult to reason about types at a given call site as some unexpected
conversion could have happened. An example is the, now deprecated, \Java
collection conversion. In an earlier iteration, \Scala defined implicit
conversions between \Java collections and its own, such as:
\begin{lstlisting}
  implicit def asJavaCollection[A](it: Iterable[A]): java.util.Collection[A]
  implicit def collectionAsScalaIterable[A](i: java.util.Collection[A]): Iterable[A]
\end{lstlisting}
As they were often imported together using a wildcard import
\c{collection.JavaConversions._}, it was easy to mistakenly invoke a \Java
method on a \Scala collection and vice-versa silently converting the
collections from one to another. Furthermore, in this case, these
conversions also change semantics as the notion of equality in \Java
collections is different from \Scala collections (reference vs. element
equality).
Since implicit conversions can introduce some pitfalls, the compiler issues a
warning when compiling an implicit conversion definition. It can be suppressed
by an import (or a compiler flag) which is usually automatically done by an IDE
and thus diminishing the utility of these warnings.

\subsection{Complexity}

Implicits help programmers by hiding the ``boring parts'' of programming
tasks, the plumbing that does not require skill or attention.  The problem
is that, as the above idioms demonstrate, implicits are also used for subtle
tasks. Their benefits can turn into drawbacks.  One way to measure the
potential complexity of implicits is to look at the work done by the
compiler. When implicits work, programmers need not notice their
presence. But when an error occurs, the programmer suddenly has to
understand the code added by \Scalac. For example, a comparison of two
tuples \c{(0,1)<(1,2)} gets expanded to
\c{orderingToOrdered((0,1))(Tuple2(Int, Int))<(1,2)}. The compiler injects
two additional calls (\c{orderingToOrdered} implicit conversion, \c{Tuple2}
type class) with two implicit arguments (\c{Int}). The question is how much
of this \emph{filling} there is.

Tooling can help navigate the complexity added by implicits. The plugin for
IntelliJ IDEA has a feature that can show implicit hints, including the
implicit resolution in the code editor.  This effectively reveals the
injected code making it an indispensable tool for debugging. However,
turning the implicit hints on severely hinders the editor performance,
creating a significant lag when working with implicits-heavy files.  The
second problem with this is that the IntelliJ compiler is not the same as
\Scalac, and often implicit resolution disagrees between the two compiler
implementations.

Another common problem that hinders understanding is related to implicit
resolution. Eligible implicits for both conversions and parameters are searched
in two different scopes. The search starts in
the lexical scope that includes local names, enclosing members and imported
members and continues in the implicit scope that consists of all companion
objects associated with the type of the implicit parameter. The advantage of
the implicit scope is that it does not need to be explicitly imported. This
prevents errors caused by missing imports for which, due to the lack of global
implicit coherence, the compiler cannot give a better error message than a type
mismatch, \emph{``member not found''} or \emph{``could not find implicit
  value''}. Implicit scope has a lower priority allowing users to override
defaults by an explicitly importing implicit definition into the lexical scope.
A consequence of this is that an import statement can change program semantics.
For example, in the code bellow contains two late trait implementations of a
trait \c{T} for a class \c{A}: \c{C1} defined in implicit scope of the class
\c{A}, and \c{C2} defined in an unrelated object \c{O}:
\begin{lstlisting}
  trait T { def f: Int }
  class A; object A { implicit class C1(a: A) extends T { def f = 1 } }
  object O { implicit class C2(a: A) extends T { def f = 2 } }
  new A().f
\end{lstlisting}
At the call to \c{f}, the compiler will use the \c{C1} conversion resolved from
the implicit scope so the result will be \c{1}. However, if later there is an
\c{import O._} before the call site, the expression will return \c{2}. The
import will bring \c{C2} into the lexical scope prioritizing \c{C2} over
\c{C1}.

Further, implicits defined in the lexical scope follow the name binding rules
and thus can be shadowed by explicit local definitions. For instance, adding
any definition with a name \c{C2} (\Eg \c{val C2 = null}) into the scope before
the call to \c{f} will result in returning again \c{1}, since the imported
\c{O.C2} implicit will be shadowed by this local definition. In the case \c{C1}
did not exits, the compiler will simply emit \emph{``value f is not a member of
  A''} error. To avoid this, library authors try to obfuscate the implicits
names which in turn affects the ergonomics. A notable example is in the \Scala
standard library where the implicit providing a proof that two types are in a
sub-type relationship is named \c{\$conforms} in order to prevent a potential
shadowing with a locally defined \c{conforms} method.\footnote{Reported in
  Scala issue \#7788, \Cf \url{https://github.com/scala/bug/issues/7788}}

\subsection{Overheads}

Implicit resolution together with macro expansion can sometimes significantly
increase compilation time. To illustrate the problem, consider the JSON
serialization of algebraic data types using the \Circe\footnote{\Cf
  \url{https://github.com/circe/circe}}, a popular JSON serialization library.
We define two ADTs: \c{case class A(x: String)} and \c{case class B(xs:
  List[A], ys: List[A])}, and a method to print out their JSON representation:
\begin{lstlisting}
def print(a: A, b: B) = println(a.asJson, b.asJson)
\end{lstlisting}
The \c{asJson} method is an extension method defined in the \Circe as \c{def
  asJson(implicit encoder: Encoder[T]): Json}. It uses an implicit parameter of
type \c{Encoder[T]} effectively limiting its applicability to instances that
define corresponding encoder. For the code to compile, two encoders
\c{Encoder[A]} and \c{Encode[B]} that turn \c{A} and \c{B} into \c{Json} are
needed. The \Circe library gives three options for creating the encoder:
manual, semi automated and automated.

\begin{figure}[!h]
\begin{minipage}[t]{.51\textwidth}
\begin{lstlisting}
object manual {
  implicit val eA: Encoder[A] = (a: A) =>
   obj("x"->str(a.x))
  implicit val eB: Encoder[B] = (b: B) =>
   obj("xs" -> arr(b.xs.map(_.asJson)),
       "ys" -> arr(b.ys.map(_.asJson)))
}
\end{lstlisting}
\centering \textsf{(a) manual}
\end{minipage}\hfill
\begin{minipage}[t]{.49\textwidth}
\begin{lstlisting}

object semiauto {
  import io.circe.generic.semiauto._

  implicit val eA: Encoder[A] = deriveEncoder[A]
  implicit val eB: Encoder[B] = deriveEncoder[B]
}
\end{lstlisting}
\centering \textsf{(b) semi-automated}
\end{minipage}

\begin{tabular}{c}
\begin{lstlisting}
import io.circe.generic.auto._
\end{lstlisting} \\
\textsf{(c) automated}
\end{tabular}
\caption{Type class derivations}\label{lst:derivation}
\end{figure}

\noindent
The manual encoding involves implementing the single method in \c{Encoder},
manually creating an instance of \c{Json} with the appropriate fields
(\Cf~Listing.~\ref{lst:derivation}a). While simple, it is a boilerplate code.
The semi-automated solution delegates to \c{derivedEncored} that synthesizes
the appropriate type at compile time through implicit type class derivation and
macros (\Cf~Listing.~\ref{lst:derivation}b). The fully automated solution, does
not require extra code at the client side beside importing its machinery
(\Cf~Listing.~\ref{lst:derivation}c). Compile time is affected by the choice of
approach; taking the manual as a base line, semi-automated is 2.5x slower and
automated is 3.8x slower.

\begin{wraptable}{r}{.4\columnwidth} \center
  \vspace{-2mm}
  \caption{Count of implicit resolutions and macro expansions, and timing
      of the typer phase in \Scalac 2.12.8 with \c{-Ystatistics:typer}
      flag.} \label{tab:typer}
  \begin{tabular}{l@{}rrr@{}}
        \hline
        ~       & Implicits    & Macros      & Time \\ \hline\hline
        Manual  & 13           & 0           & .1s \\
        Semi    & 35           & 51          & .3s \\
        Auto    & 52           & 78          & .5s \\
        \hline
    \end{tabular}
\end{wraptable}

The reason for this compile-time slow-down is the increase in the number of
implicit resolutions triggered and macro expansion as shown in
Table~\ref{tab:typer}. The difference between the automated and semi-automated
is that the former \emph{caches} the derived instances in the implicit values
\c{eA} and \c{eB} and so the \c{eA} which is synthesized before \c{eB} will be
reused for deriving \c{eB}. The automated derivation synthesizes new instances
for each application. In this simple example, it generates 140 additional lines
of code at the \c{println}\footnote{Measured in the expanded code obtained from
  \c{-Xprint:typer} compiler flag}. Caching of derived type classes was already
reported to significantly improve the compilation time of various
projects~\cite{torreborre17,canteor18_scalac_profiling}. One difficulty is that
since the implicit scope is \emph{invisible}, it is harder to figure out which
implicits are derived where and which are causing slowdowns. Currently, the
only way is to use a \emph{scalac-profiling}\footnote{\Cf
  \url{https://github.com/scalacenter/scalac-profiling}}, compiler plugin which
outputs more detailed statistics about implicit resolution and macro expansion.

\section{Scala Analysis Pipeline}\label{sec:analysis}

We have implemented a data analysis pipeline targeting large-scale analysis
of \Scala programs. To the best of our knowledge, this is the only pipeline
able to scale to thousands of projects. Our infrastructure can be extended
for other analyses and it is available in open source.

\begin{figure}[t!]
  \vspace{2mm}
  \includegraphics[width=\textwidth]{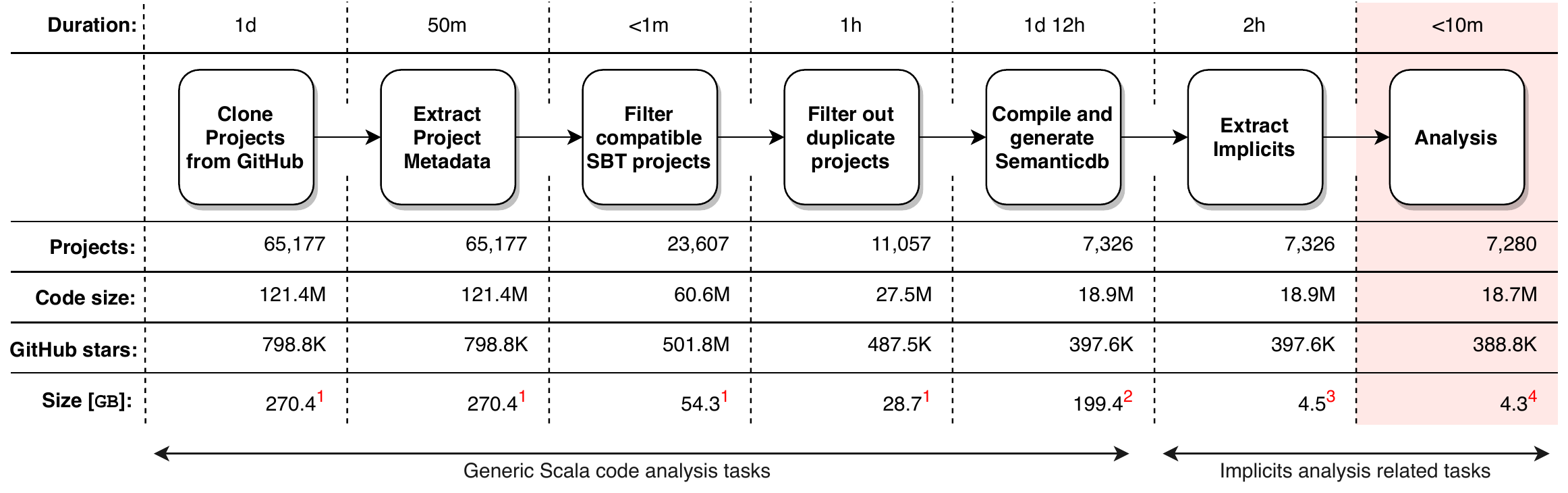}
  \caption{Scala Analysis pipeline. (1) is the size of source code, (2) is
    the size of source plus compiled code and generated \SDB, (3) is the
    size of extracted implicits data model, (4) is the size of exported CSV
    files. The code size include tests.}
    \label{fig:pipeline}
\end{figure}

Figure~\ref{fig:pipeline} gives an overview of the pipeline; every step
shown in the figure is fully automated. The first step is to download
projects hosted on \GH. Next, gather basic metadata and in particular
infer the build system each project uses.  Incompatible projects are
discarded in the next step. These are projects that do not meet the
technical requirements of the analysis tools.  The fourth step is to use the
DéjàVu tool~\cite{lopes17} to filter out duplicate projects. The fifth step is
to attempt to compile the corpus and generate semantic information.  The
final step is to load the extracted data and analyze them.  The pipeline is
run in parallel using GNU parallel~\cite{tange2011_parallel} but the
analysis is resource intensive. On our server (Intel Xeon 6140, 2.30GHz with
72 cores and 256GB of RAM) we were not able to compile more than 12 projects
in parallel.

The pipeline is reusable for other semantic analyses on Scala code bases, as only
the last two steps relate specifically to implicits. At the end of the
\emph{Compile and generate \SDB} task, the corpus contains built projects with
extracted metadata and \SDB files---these \SDB files contain syntactic
information as well as semantic information (Scala symbols and types).

The pipeline logs all the steps for each
project and provide an aggregated summaries.  The analysis is done in R, and
even though it is possible to load Google Protocol Buffers into R, it is not
practical. Thus, we first aggregate the extracted data and export them into CSV
format, which is more natural to work with in R. This is implemented in
\textasciitilde{}500 lines of make files and \textasciitilde{}5K of R code. The
implicit extractor is written in \textasciitilde{}7.2K lines of \Scala code.

The pipeline uses \ScalaMeta\footnote{\Cf \url{https://scalameta.org/}}, a library
that provides a high-level API for analyzing programs. One part of this library
is a compiler plugin that for each compilation unit produces a data model with
syntactic and semantic information. This includes a list of defined and
referenced symbols as well as synthetic call sites and parameters injected by
the compiler. The result is stored in a binary
\SDB\footnote{\Cf \url{https://scalameta.org/docs/semanticdb/specification.html}}
file (in Google Protocol Buffer serialization format). It can also extract
symbol information from compiled classes allowing us to find implicits defined
in external project dependencies. Note that this analysis would have not been
possible with only syntactic information; compile-time information like types is
required to match up call site and declaration site due to the fact that
implicits themselves are type-directed rewritings performed by the compiler at
type-checking time.

Based on this we have built a tool that extract implicit declarations and call
sites. There are two limitations with \ScalaMeta: it is limited to certain
versions of \Scala (2.11.11 in the 2.11 branch and 2.12.4 in the 2.12 branch),
and it does not support \emph{white-box macros} (\Ie macros without precise
signatures in the type system before their expansion)~\cite{burmako17}.

Another thing to consider when using \SDB is that it requires compiling the
projects. The \Scala compiler is about an order of magnitude slower than a Java
compiler\footnote{\Cf \url{https://stackoverflow.com/a/3612212/219584}} and the
\SDB compiler plugin adds additional overhead. For our analysis \SBT is used to
rebuild each project three times. There is no easy way around this. As noted
above, lightweight, syntax-based approaches using regular expressions or
pattern matching over AST nodes would not work because the call sites that use
implicits are not visible in the source/AST, and to identify these patterns
requires resolving terms and types from the declaration- and use-sites.

\Scala projects are compiled by build tools which are responsible for
resolving external dependencies. We chose \SBT as it is the most-used tool
in the \Scala world. Since version 0.13.5 (August 2014), it supports custom
plugins which we use to build an extractor of metadata. Next to the version
information and source folder identification, the extracted metadata gives
us information about project internal and external dependencies. This is
necessary for assembling project's class-path that is used to resolve symbols
defined outside of the project.

\subsection{Implicit Extraction}\label{sec:implicit-extraction}

The \SDB model contains low-level semantic information about each
compilation unit. This includes synthetics, trees added by compilers that do
not appear in the original source (\Eg inferred type arguments,
for-comprehension desugarings, \c{C(...)} to \c{C.apply(...)} desugarings,
implicit parameters and call sites). These trees are defined as
transformations of pieces of the original \Scala AST and as such they use
quotes of the original sources. For example, the following \Scala code:
\begin{lstlisting}
import ExecutionContext.global; Future(1)
\end{lstlisting}
will have two synthetic trees injected by the compiler:
\begin{lstlisting}
- ApplyTree(OriginalTree(1,60,1,86), IdTree("EC.global"))
- TypeApplyTree(SelectTree(OriginalTree(1,60,1,83),IdTree("Future.apply()")),TypeRef("Int"))
\end{lstlisting}
In this form, \SDB is not convenient for higher-level queries about the use of
implicits. In order to do this, we transform \SDB into our own model that has
declarations and call sites resolved. This is done in two steps. First, we
extract implicit declarations by traversing each compilation unit and
collecting declarations with the \c{implicit} modifier. For each declaration,
we resolve its type using the symbol information from the \SDB and the project
class path. This is done recursively in the case the declaration type has
parents. Next, we look into the synthetic trees and extract inserted implicit
function applications. Together with the project metadata, both declaration and
call sites are stored in a tree-like structure using the Google Protocol Buffer
format. In our example, the extractor will produce 13 declarations and one
implicit call site including:
\begin{lstlisting}
  // def apply[T](body: => T)(implicit executor: EC)
- Declaration("Future.apply()", DEF, ret=Ref("Future.apply().[T]"), params=List(
    ParamList(Param("body", Ref("Future.apply().[T]"))),
    ParamList(Param("executor", Ref("EC"), isImplicit=true))))
  // implicit val global: EC
- Declaration("EC.global", VAL, ret=Ref("EC",List()), isImplicit=true)
  // Future.apply[Int](1)(EC.global)
- CallSite("Future.apply()", typeArgs=Ref("Int"), implicitArgs=Ref("EC.global"))
\end{lstlisting}
Such model can be queried using the standard \Scala collection API. For example, we can list a project's
\c{ExecutionContext} declarations and the corresponding call sites that use them as follows:
\begin{lstlisting}
val declarations = proj.declarations filter (dcl =>
  dcl.isImplicit && dcl.isVal && dcl.returnType.isKindOf("EC"))
val callsites = {
  val ids = declarations.map(_.declarationId).toSet
  proj.implicitCallsites filter (cs =>
    cs.implicitArguments exists (arg => ids contains arg))
}
\end{lstlisting}
The extractor is run per project in parallel and the results are merged into
one binary file. This file can be streamed into a number of processors that
export information about declarations, call sites, implicit conversions and
implicit parameters into CSV files.

\section{Project Corpus}\label{sec:corpus}

For this paper we analyzed \TgCorpus projects consisting of \TgCorpusCodeRnd
lines of \Scala code (including \TgCorpusTestCodeRnd lines of tests and
\TgCorpusGencodeRnd lines of generated code). Most projects are small, the
median is \TgCorpusCodeMedian lines of code, but the corpus also includes
projects with over 100K lines of source code. \TgCorpusScalaEleven projects
use \Scala 2.11 but they account for less code (\TgCorpusScalaElevenCodePct)
and fewer stars (\TgCorpusScalaElevenStarsPct). For the remainder of the
paper we partition our corpus in four categories: {\bf small apps}
are project with fewer than 1,000 LOC, {\bf large apps} are
projects with more than 1,000 LOC, {\bf libraries} are projects
that are listed on \Scaladex. We also extract the test code from all projects
into the {\bf tests} category.  \Scaladex is a package index of
projects published in Maven Central and Bintray repositories.  These labels
are somewhat ad-hoc as there is not always a strong reason behind the
addition of a project to Maven Central or Bintray.  However, manual
inspection suggests that most of the projects that appear on \Scaladex are
intended for reuse.

\begin{table}[!h]
  \caption{Project categories}\label{tab:corpus-categories}
  \centering \small
  \begin{tabular}{lllll}  \hline
    Category    & Projects & Code size      & \GH stars       & Commits         \\ \hline
    Small apps. & 3.3K     & 1M (mean=0.3K) & 28K (mean=8)    & 139K (mean=41)  \\
    Large apps. & 1.3K     & 5M (mean=4.0K) & 74K (mean=57)   & 425K (mean=325) \\
    Libraries   & 2.6K     & 6M (mean=2.4K) & 285K (mean=108) & 712K (mean=271) \\
    Tests       & 5.4K     & 5M (mean=1.1K) & -               & -               \\ \hline
  \end{tabular}
\end{table}

\begin{figure}[!h]  \centering
  \includegraphics[width=1.01\linewidth]{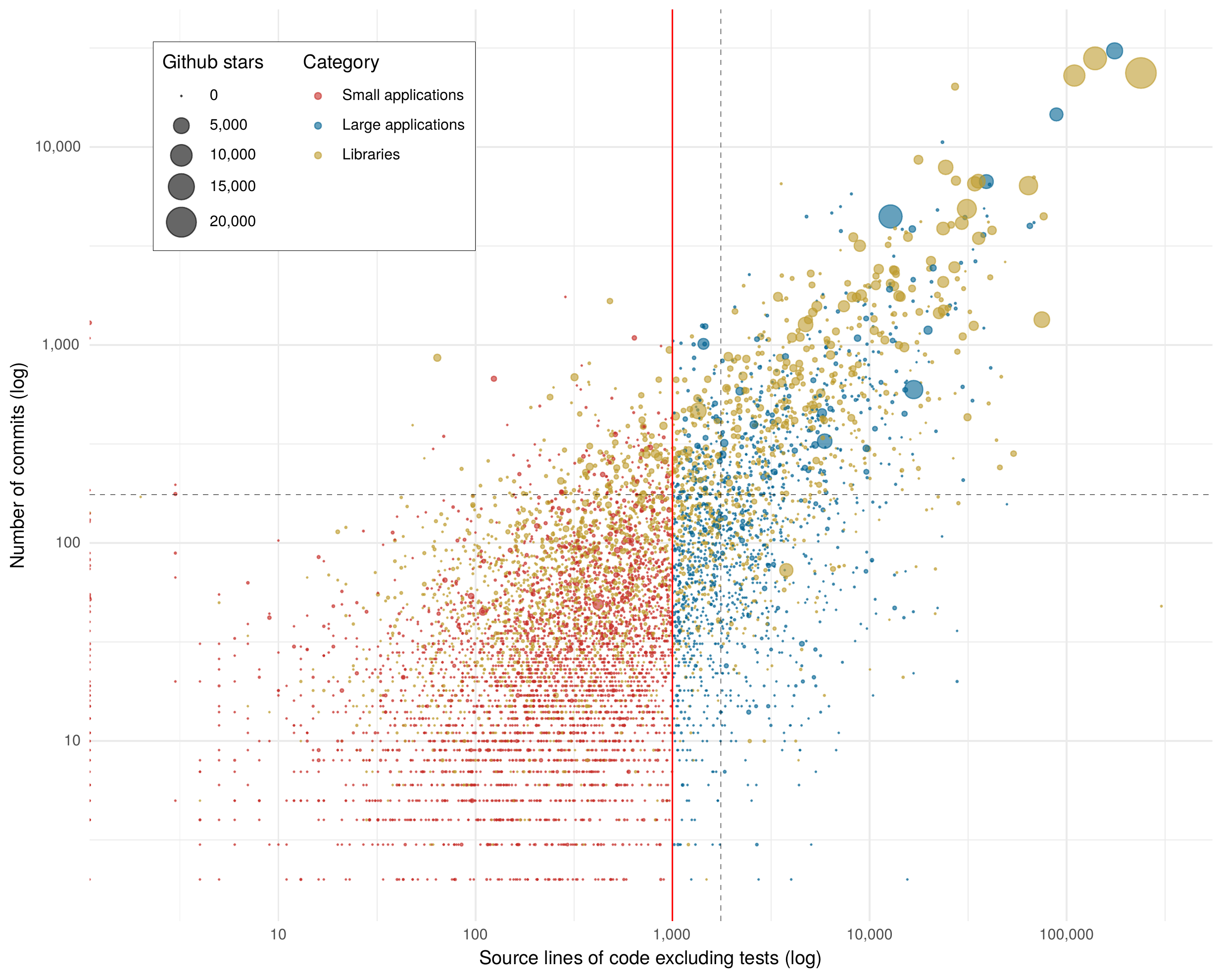}
  \caption{Corpus overview}\label{fig:corpus}
\end{figure}

\noindent
Figure~\ref{fig:corpus} shows all projects, the size of the dots reflects
number of stars, the color their category (large/small apps or libraries), the
x-axis indicates the number of lines of code (excluding \TgCorpusTestCodeRnd
lines of tests) in log scale, the y-axis gives the number of commits to the
project in log scale. Solid lines indicate the separation between small
and large applications. Dotted lines indicate means.

The corpus was obtained from publicly available projects listed in the
GHTorrent database~\cite{gousi13} and \Scaladex. The data was downloaded between
January and March 2019. We started with \TgDownloadedProjects non-empty,
non-fork projects, which together contained \TgDownloadedProjectsCodeRnd lines
of code. We filtered out projects that were not compatible with our analysis
pipeline (e.g., projects using early versions of Scala) and removed duplicates.
\TgSbtProjectsRnd use \SBT as their build system (other popular build systems
are Maven with \TgMavenProjectsRnd projects and Graddle with
\TgGraddleProjectsRnd). From the \SBT projects, \TgCompatibleSbtProjectsRnd use
\SBT version $0.13.5+$ or $1.0.0+$ that is required by our analysis. We thus
discarded about half of the downloaded code.

\begin{table}[b!]
  \caption{Top 40 open source projects}\label{tab:corpus-sample}
  \centering \footnotesize
\begin{tabular}{lrrrrcc}
  \hline
Project & GitHub stars & Code size & Commits & Duplication & Scala version & Scaladex \\ 
  \hline
\href{https://github.com/apache/spark/}{apache/spark} & 21,067 & 238,062 & 23,668 & 0.4 & 2.12.8 & Y \\ 
  \href{https://github.com/apache/predictionio/}{apache/predictionio} & 11,696 & 12,764 & 4,461 & 0 & 2.11.12 & N \\ 
  \href{https://github.com/scala/scala/}{scala/scala} & 11,386 & 139,300 & 28,062 & 0.9 & 2.12.5 & Y \\ 
  \href{https://github.com/akka/akka/}{akka/akka} & 9,666 & 109,359 & 22,966 & 0.001 & 2.12.8 & Y \\ 
  \href{https://github.com/gitbucket/gitbucket/}{gitbucket/gitbucket} & 7,612 & 31,144 & 4,874 & 0 & 2.12.8 & Y \\ 
  \href{https://github.com/twitter/finagle/}{twitter/finagle} & 7,003 & 63,976 & 6,386 & 0.01 & 2.12.7 & Y \\ 
  \href{https://github.com/yahoo/kafka-manager/}{yahoo/kafka-manager} & 6,958 & 16,733 & 596 & 0.5 & 2.11.8 & N \\ 
  \href{https://github.com/ornicar/lila/}{ornicar/lila} & 5,218 & 175,054 & 30,617 & 0.01 & 2.11.12 & N \\ 
  \href{https://github.com/rtyley/bfg-repo-cleaner/}{rtyley/bfg-repo-cleaner} & 5,014 & 1,351 & 465 & 0 & 2.12.4 & Y \\ 
  \href{https://github.com/linkerd/linkerd/}{linkerd/linkerd} & 4,910 & 74,775 & 1,344 & 0.003 & 2.12.1 & Y \\ 
  \href{https://github.com/fpinscala/fpinscala/}{fpinscala/fpinscala} & 4,244 & 5,914 & 327 & 1 & 2.12.1 & N \\ 
  \href{https://github.com/haifengl/smile/}{haifengl/smile} & 4,242 & 4,731 & 1,271 & 0 & 2.12.6 & Y \\ 
  \href{https://github.com/gatling/gatling/}{gatling/gatling} & 4,151 & 24,322 & 7,900 & 0 & 2.12.8 & Y \\ 
  \href{https://github.com/scalaz/scalaz/}{scalaz/scalaz} & 4,079 & 34,146 & 6,523 & 0 & 2.12.8 & Y \\ 
  \href{https://github.com/mesosphere/marathon/}{mesosphere/marathon} & 3,823 & 39,097 & 6,694 & 0.03 & 2.12.7 & N \\ 
  \href{https://github.com/sbt/sbt/}{sbt/sbt} & 3,782 & 35,574 & 6,726 & 0.4 & 2.12.8 & Y \\ 
  \href{https://github.com/twitter/diffy/}{twitter/diffy} & 3,375 & 3,778 & 73 & 0 & 2.11.7 & Y \\ 
  \href{https://github.com/lampepfl/dotty/}{lampepfl/dotty} & 3,278 & 88,680 & 14,616 & 0.3 & 2.12.8 & N \\ 
  \href{https://github.com/twitter/scalding/}{twitter/scalding} & 3,113 & 29,346 & 4,133 & 0 & 2.11.12 & Y \\ 
  \href{https://github.com/typelevel/cats/}{typelevel/cats} & 3,093 & 23,607 & 3,878 & 0.009 & 2.12.7 & Y \\ 
  \href{https://github.com/scalanlp/breeze/}{scalanlp/breeze} & 2,816 & 35,747 & 3,461 & 0.002 & 2.12.1 & Y \\ 
  \href{https://github.com/scalatra/scalatra/}{scalatra/scalatra} & 2,382 & 8,914 & 3,174 & 0.3 & 2.12.8 & Y \\ 
  \href{https://github.com/netflix/atlas/}{netflix/atlas} & 2,288 & 22,474 & 1,450 & 0 & 2.12.8 & Y \\ 
  \href{https://github.com/spark-jobserver/spark-jobserver/}{spark-jobserver/spark-jobserver} & 2,286 & 7,403 & 1,571 & 0.3 & 2.11.8 & Y \\ 
  \href{https://github.com/twitter/util/}{twitter/util} & 2,243 & 26,927 & 2,472 & 0.2 & 2.12.7 & Y \\ 
  \href{https://github.com/slick/slick/}{slick/slick} & 2,188 & 23,622 & 2,084 & 0 & 2.11.12 & Y \\ 
  \href{https://github.com/laurilehmijoki/s3\_website/}{laurilehmijoki/s3\_website} & 2,178 & 1,435 & 1,014 & 0 & 2.11.7 & N \\ 
  \href{https://github.com/twitter/summingbird/}{twitter/summingbird} & 2,011 & 9,057 & 1,790 & 0.3 & 2.11.12 & Y \\ 
  \href{https://github.com/MojoJolo/textteaser/}{MojoJolo/textteaser} & 1,942 & 420 & 49 & 0 & 2.11.2 & N \\ 
  \href{https://github.com/twitter/finatra/}{twitter/finatra} & 1,888 & 14,071 & 1,772 & 0.001 & 2.12.6 & Y \\ 
  \href{https://github.com/twitter/algebird/}{twitter/algebird} & 1,836 & 23,676 & 1,502 & 0 & 2.11.12 & Y \\ 
  \href{https://github.com/scala-exercises/scala-exercises/}{scala-exercises/scala-exercises} & 1,775 & 5,398 & 1,570 & 0 & 2.11.11 & Y \\ 
  \href{https://github.com/circe/circe/}{circe/circe} & 1,633 & 8,140 & 1,749 & 0.006 & 2.12.8 & Y \\ 
  \href{https://github.com/datastax/spark-cassandra-connector/}{datastax/spark-cassandra-connector} & 1,569 & 11,120 & 2,418 & 0.2 & 2.11.12 & Y \\ 
  \href{https://github.com/rickynils/scalacheck/}{rickynils/scalacheck} & 1,480 & 4,038 & 1,091 & 0 & 2.12.6 & Y \\ 
  \href{https://github.com/monix/monix/}{monix/monix} & 1,466 & 33,749 & 1,251 & 0 & 2.12.8 & Y \\ 
  \href{https://github.com/http4s/http4s/}{http4s/http4s} & 1,459 & 27,412 & 6,765 & 0.003 & 2.12.7 & Y \\ 
  \href{https://github.com/sangria-graphql/sangria/}{sangria-graphql/sangria} & 1,442 & 14,999 & 975 & 0.2 & 2.12.7 & Y \\ 
  \href{https://github.com/spotify/scio/}{spotify/scio} & 1,439 & 20,477 & 2,659 & 0.002 & 2.12.8 & Y \\ 
  \href{https://github.com/coursier/coursier/}{coursier/coursier} & 1,417 & 13,313 & 1,984 & 0 & 2.12.8 & Y \\ 
   \hline
\end{tabular}

\end{table}

For duplicates, the problem is that even without \GH forks, the corpus still
contained unofficial forks, \Ie copies of source code. For example, there
were \TgSparkProjectsBefore copies of \Spark. Since \Spark is the largest
\Scala project (over 100K LOC), keeping them would significantly skew the
subsequent analysis as \TgSparkProjectsBeforeCodePct of the entire data set
would be identical. In general, getting rid of duplicate projects is
difficult task as one needs to determine the origins of individual files. We
use that following criteria to retain a project:
\begin{inparaenum}[(1)]
\item it must have more than one commit,
\item it must be active for at least 2 months,
\item it must be in \Scaladex or have less than 75\% of file-level
  duplication or more than 5 stars on \GH, and
\item it must be in \Scaladex or have less than 80\% duplication or more than
  500 stars on \GH.
\end{inparaenum}
These rules were tuned to discard as many duplicates as possible while
keeping originals.  While large numbers of \GH stars do not necessarily mean
that a project widely-used, originals tend to have higher star counts than
copies. The actual thresholds were chosen experimentally to make sure we
keep all the bigger (> 50K LOC) popular \Scala projects without any
duplicates.  We excluded \TgDuplicated projects (\TgDuplicatedCodeRnd lines
of code). While this is over half of the source code from the compatible
\SBT projects, we lost fewer than \TgLostStarsInDuplicatedProjectsPct stars.

From the resulting \TgSelectedProjects projects, we were able to
successfully compile \TgExtractedProjects projects. \TgFailedProjectsAll
projects failed to build. We follow the standard procedure of building \SBT
projects. If a project required additional steps, we marked it as
failed. The following are the main sources of failures:
\begin{compactitem}[$-$]
\item \emph{Missing dependencies}
  (\TgFailedProjectsMissingDependenciesRnd). Most missed dependencies were
  for \ScalaJS (\TgFailedProjectsMissingScalajs), a \Scala-to-JavaScript
  compiler with a version that was likely removed because of security
  vulnerabilities. The next most frequent issue was due to snapshot versions
  (\TgFailedProjectsMissingSnapshots) that were no longer available. The
  remainder were libraries that were taken down or that reside in
  non-standard repositories. Following common practice, we use a local proxy
  that resolves dependencies. No additional resolvers were configured. The
  proxy downloaded 204K artifacts (110GB).
\item \emph{Compilation error} (\TgFailedProjectsCompileError). Some commits
  do not compile, and others fail to compile due our restriction on \Scala
  versions.  \ScalaMeta requires \Scala $2.11.9+$ or $2.12.4+$. Some
  projects are sensitive even down to the path version number. Some of these
  version upgrades might have also caused the missing dependencies in case
  the required artifact was built for a particular \Scala version.
\item \emph{Broken build} (\TgFailedProjectsBrokenBuild). The \SBT could not
  even start due to errors in the \c{build.sbt}.
\item \emph{Empty build} (\TgFailedProjectsEmptyBuild). Running \SBT did not
  produce class files, leaving the projects \emph{empty}. This happens when
  the build has some non-standard structure.
\end{compactitem}

\noindent
Finally, in the analysis, we discarded \TgDataCleanupRemovedProjects
projects (\TgDataCleanupProjectsLostCodePct of the code) because some of
their referenced declarations were not resolvable (the \ScalaMeta symbol
table did not return any path entry) and inconsistencies in \SDB.
Table~\ref{tab:corpus-sample} lists some of the top rated projects that were
included in the final corpus, including number of stars, lines of code,
number of commits, level of duplication, \Scala version and whether it is
listed in \Scaladex.

\newpage
\section{Analyzing Implicits Usage}

This section presents the results of our analysis and paints a picture of the
usage of implicits in our corpus of \Scala programs. We follow the structure of
Section~\ref{sec:scala-implicits} and give quantitative data on the various
patterns and idioms we presented including details about how identified them. We
further discuss the impact of implicits on code comprehension and compilation
time.

Identifying implicits requires performing a number of queries on the data files
produced by our pipeline. Doing this also turned out to be necessary to remove
duplication due to compilation artifacts. These come from projects compiled for
multiple platforms and projects compiled for multiple major versions of \Scala.
While the main compilation target for \Scala projects is Java byte-code
(\TgProjectsTargetingJvm projects), JavaScript and native code are also
potential targets. To prevent double counting, we make sure that shared code is
not duplicated. Since \Scala 2.11 and 2.12 are not binary compatible, libraries
supporting both branches cross compile to both versions. We take care to
compile only to one version.

In the remainder of this paper, when we refer to the ``\Scala library,''
``\Scala standard library,'' or sometimes just to ``\Scala'' we mean code
defined in \c{org.scala-lang:scala-library} artifact.

\paragraph{Overview of Results}

\begin{figure}[!b]
  \includegraphics[width=.8\linewidth]{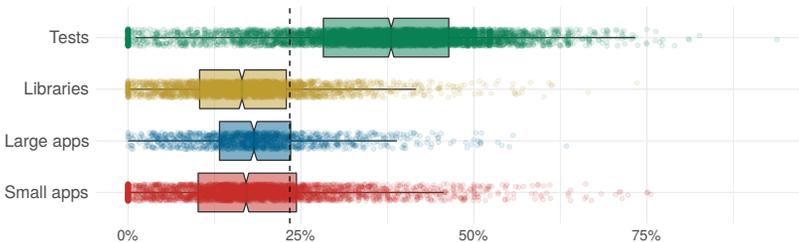}
  \caption{Ratio of implicit call}\label{cs}
\end{figure}

Out of the \TgCorpus analyzed projects, \TgProjectsUsingImplicits
(\TgProjectsUsingImplicitsPct) have at least one implicit call site. From over
\TgCallSitesRnd call sites in the corpus (explicit and implicit combined),
\TgImplicitCallSitesRnd are call sites involving implicits. Most of these calls
are related to the use of implicit parameters (\TgIpInCallSitesPct).
Figure~\ref{cs} shows for each category a distribution of implicit call site
ratios. The box is the 25th/75th percentiles and the line inside the box
represents the median with the added jitters showing the actual distribution.
For applications and libraries, the median is similar. It is smaller
\textasciitilde{}\TgImplicitsMainCallsitesPctMedian. In the case of test code,
it is more than double, \TgImplicitsTestCallsitesPctMedian. There tend to be
more implicit call sites in tests than in the rest of the code. That is not
surprising because the most popular testing frameworks heavily rely on
implicits. Across the project categories the median is
\TgImplicitsCallsitesPctMedian (shown by the dashed line)---\Ie~\emph{one out
  of every four call sites involves implicits}.

\begin{figure}[!b]
  \includegraphics[width=.8\linewidth]{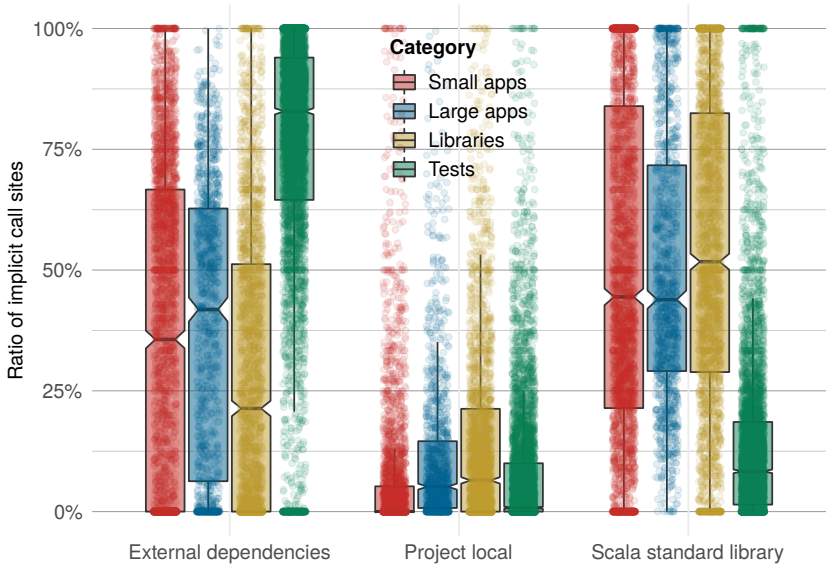}
  \caption{Origin of parameter declarations}\label{fig:callsites-locations}
\end{figure}

Figure~\ref{fig:callsites-locations} shows the distribution of the declarations
that are being called from the implicit call sites. There is a big difference
between the test and non-test category. In the case of the both applications
and libraries, most implicits used come from the standard library, followed by
their external dependencies. The main sources of implicits in \Scala are
collections, concurrency and reflection packages together with the omnipresent
\c{scala.Predef} object.

The collections are used by \TgIpScalaCollectionAllProjectsUsingPct projects
(from \TgIpScalaCollectionAppBigProjectsUsingPct in large apps to
\TgIpScalaCollectionTestProjectsUsingPct in tests). Most of the collection
transforming operations such as \c{map}, use a builder factory passed as an
implicit parameter \c{CanBuildFrom}. \TgIpScalaCollectionLibCallsitesPct of all
implicit call sites involving methods that use this implicit parameter appear
in libraries.
Implicit parameters are used for reflection. Instances of \c{Manifest},
\c{ClassTag} or \c{TypeTag} classes can be requested from the compiler to be
passed as implicit arguments, allowing one to get in-depth information about
the type parameters of a method at runtime, circumventing the limitation of
Java's type erasure. This is used a lot in large applications
(\TgIpScalaReflectionAppBigProjectsUsingPct). Less in libraries
(\TgIpScalaReflectionLibProjectsUsingPct) and small projects
(\TgIpScalaReflectionAppSmallProjectsUsingPct) or tests
(\TgIpScalaReflectionTestProjectsUsingPct).
Over half of all the large projects
(\TgIpScalaConcurrencyAppBigProjectsUsingPct) and third of libraries
(\TgIpScalaConcurrencyAppSmallProjectsUsingPct) employ some concurrency
routines from the \Scala standard library.
\c{scala.Predef} defines basic conversion like \c{String} to \c{StringOps}
(extending the functionality of Java strings) or an arrow association, allowing
one to use \c{a->b} to create a tuple of \c{(a,b)}. These are used by
almost all the projects regardless of category.

Excluding the \Scala standard library and testing frameworks, the rest of the
implicits in the case of application and libraries come from a number of
different external dependencies. There are some well known and projects with
rich set of implicit usage such as the Lightbend/Typesafe stack with Play (a
web-application framework, used in \TgCallsitesFormPlay of implicit call
sites), Slick (object-relational mapping, \TgCallsitesFormSlick) or \Akka (an
actor framework, \TgCallsitesFormAkka). These libraries define domain-specific
languages which, in order to fit well in the host language yet to appear to introduce
different syntactic forms, heavily rely on implicits. Next to a more flexible syntax (as
compared to Java or C\#), implicits are the main feature for embedding DSLs.

In the case of tests, the vast majority of implicits comes from project
dependencies, which are dominated by one of the popular testing frameworks.
These frameworks define DSLs in one form or another, striving to provide an API
that reads like English sentences. For example a simple test:
\begin{lstlisting}
"Monte Carlo method" should "estimate pi" in { MCarloPi(tries=100).estimate === 3.14 +- 0.01 }
\end{lstlisting}
contains six implicit call sites. Four are implicit conversions adding methods
\c{should} to \c{String}, \c{in} to \c{ResultOfStringPassedToVerb} (the
resulting type of calling the \c{should} method), \c{===} and \c{+-} to
\c{Double}. Three of them additionally take implicit parameters for
pretty-printing, source position (generated by a macro), test registration, and
floating point operations. The implicit macro generating the source position is
actually the single most used implicit parameter in the corpus with
\TgUseScalacticSourcePositionRnd instances. Excluding the test frameworks, the
ratio of implicit locations become very close to that of the main code, with
collections and \c{scala.Predef} dominating the distribution.

\subsection{Implicit Conversion}\label{sec:analysis-implicit-conversion}

We recognize conversions by finding signatures that are either:
\begin{inparaenum}
\item an \c{implicit def} with one non-implicit parameter (and 0+ implicit
  parameters) and a non-\c{Unit} return, or
\item an implicit \c{val}, \c{var} or \c{object} that extends a function
  type \c{T=>R} such that \c{R} is not \c{Unit}.
\end{inparaenum}
Note, that \c{implicit class} declarations are already de-sugared into a
class and a corresponding \c{implicit def}.

Table~\ref{tab:implicit-overview} summarizes conversions across the four
categories of projects; $X\ (Y\%\ Z\%)$ are such that $X$ is the number of
occurrences, $Y\%$ is the ratio of $X$ across all categories and $Z$ is a ratio
of projects identified in the given category. As expected, the majority of
implicit conversions (80\%) are defined in libraries (52\% of libraries define
at least one conversion) while most use is in the tests (61\% of all implicit
conversion call sites).

\begin{table}[!h]
  \caption{Conversions} \label{tab:implicit-overview}
  \centering \small
  \begin{tabular}{lrrrr}  \hline
    & Small Apps & Large Apps & Libraries & Tests \\
    \hline
    \bf Declarations & 2K (04\% 22\%) & 7K (13\% 58\%) & 49K (80\% 52\%) & 2K (03\% 11\%) \\
    \bf Call sites & 89K (04\% 88\%) & 384K (15\% 99\%) & 514K (20\% 94\%) & 1M (61\% 95\%) \\
    \hline
  \end{tabular}
\end{table}

\noindent
Table~\ref{ic} lists the projects declaring and using the most conversions;
each project's \GH name is followed by its star rating, lines of code, and
the number of occurrences. It is interesting to observe that the projects
that define the most conversions are not necessarily the ones which use the
most, as usage is likely correlated to project size.

  \begin{table}[!h]%
    \caption{Top conversions}\label{ic}%
    \centering \footnotesize%
\begin{tabular}{lrlr}
  \hline
Project & Declarations & Project  & Callsites \\ 
  \hline
\href{https://github.com/shadaj/slinky/}{shadaj/slinky} (265, 46K) & 34K & \href{https://github.com/exoego/aws-sdk-scalajs-facade/}{exoego/aws-sdk-scalajs-facade} (3, 302K) & 130K \\ 
  \href{https://github.com/pbaun/rere/}{pbaun/rere} (4, 14K) & 446 & \href{https://github.com/scalatest/scalatest/}{scalatest/scalatest} (782, 76K) & 116K \\ 
  \href{https://github.com/etorreborre/specs2/}{etorreborre/specs2} (642, 26K) & 440 & \href{https://github.com/apache/spark/}{apache/spark} (21K, 238K) & 60K \\ 
  \href{https://github.com/sisioh/aws4s/}{sisioh/aws4s} (7, 15K) & 402 & \href{https://github.com/akka/akka/}{akka/akka} (10K, 109K) & 30K \\ 
  \href{https://github.com/CommBank/grimlock/}{CommBank/grimlock} (29, 22K) & 385 & \href{https://github.com/gapt/gapt/}{gapt/gapt} (48, 68K) & 22K \\ 
  \href{https://github.com/scala/scala/}{scala/scala} (11K, 139K) & 346 & \href{https://github.com/ornicar/lila/}{ornicar/lila} (5K, 175K) & 17K \\ 
  \href{https://github.com/scalatest/scalatest/}{scalatest/scalatest} (782, 76K) & 343 & \href{https://github.com/psforever/PSF-LoginServer/}{psforever/PSF-LoginServer} (28, 41K) & 15K \\ 
  \href{https://github.com/scalan/special/}{scalan/special} (2, 33K) & 336 & \href{https://github.com/broadinstitute/cromwell/}{broadinstitute/cromwell} (384, 65K) & 15K \\ 
  \href{https://github.com/scalaz/scalaz/}{scalaz/scalaz} (4K, 34K) & 301 & \href{https://github.com/hmrc/tai-frontend/}{hmrc/tai-frontend} (0, 31K) & 14K \\ 
  \href{https://github.com/lift/framework/}{lift/framework} (1K, 42K) & 280 & \href{https://github.com/getquill/quill/}{getquill/quill} (1K, 11K) & 14K \\ 
   \hline
\end{tabular}
  \end{table}%

\noindent
Conversions are used in \TgIcAllProjectsUsingPct of all projects
(\TgIcAllProjectsUsing). There are \TgIcAllCallsitesRnd implicit conversions
or \TgIcAllCallsitesPct of all implicit call sites.  This is understandable
as it is hard to write code that does not, somehow, trigger one of the many
conversions defined in the standard library. In fact, for application code
\TgIcCallsitesFromScalaInMainPct of implicit conversions have definitions
originating in the standard library. Most conversions, \TgIcTestCallsitesPct
to be exact, happen in tests; for those, \TgIcCallsitesFromTestLibrariesPct
of them have definitions that originate from one of the two popular testing
frameworks (\ScalaTest or \Specs). If we exclude the standard library
and testing frameworks, most conversions are defined in imported code, only
about \TgIcCallsitesFromLocalDeclarationsPct are calls to conversions with
definitions local to their project.

In terms of conversion declarations, \TgIcAllProjectsDeclaringPct of projects
(\TgIcAllProjectsDeclaring) provide \TgIcAllDeclarations conversions
(\TgIcAllDeclarationsPct of all declarations) with a median of
\TgIcDeclarationsMedian per project and a s.dev of \TgIcDeclarationsSd. As
expected, testing frameworks have many declarations (\TgIcInScalatest in
\ScalaTest, \TgIcInSpec in \Specs). We note that \Slinky defines over
\TgIcInSlinkyRnd conversions (almost all programatically generated). The
reason is that this project aims at allowing one to writing React code (a
JavaScript library for building user interfaces) in \Scala in a similar manner
as to that of JavaScript. This project is hardly used, we could find only
\TgIcSlinkyUseProjects clients (with \TgIcSlinkyUseProjectsCodeRnd LOC) that
used \TgIcSlinkyUse \Slinky conversions.

The most used conversion is \c{ArrowAssoc} as it enables users to create tuples
with an arrow (\Eg \c{1 -> 2}). The next most popular is \c{augmentString},
a conversion that allows users to use index sequence methods on \c{String}
objects. On average, projects targeting JavaScript use \TgIcUsedInJsVsJvm times
more often implicit conversions than JVM projects. Most of these conversions
come from libraries that simplify front-end web development with DSLs for
recurring tasks such as DOM construction and navigation. Only
\TgIcfunAllDeclarationsRnd (\TgIcfunAllDeclarationsPct) of the implicit
conversions were defined with functional types (\Ie using \c{implicit} \c{val},
\c{var} or \c{object}); this is good as implicit values that are also
conversions can be the source of problems.

\subsection{Implicit Parameters}

We record all method and constructor declarations with implicit parameter
list.  Table~\ref{p} summarizes parameters across the four categories of
projects; $X\ (Y\%\ Z\%)$ are such that $X$ is the number of occurrences,
$Y\%$ is the ratio of $X$ over all categories and $Z$ is a ratio of projects
in the given category.

\begin{table}[!h]
  \caption{Parameters} \label{p}
  \centering \small
  \begin{tabular}{lrrrr}\hline
                     & Small Apps       & Large Apps       & Libraries         & Tests           \\ \hline
    \bf Declarations & 8K (06\% 35\%)   & 50K (32\% 73\%)  & 87K (55\% 68\%)   & 11K (07\% 23\%) \\
    \bf Call sites   & 134K (04\% 89\%) & 749K (20\% 99\%) & 691K (19\% 94\%)  & 2M (58\% 95\%)  \\ \hline
  \end{tabular}
\end{table}

\noindent
Table~\ref{ip} lists the projects declaring and using the most implicit
parameters; each project's \GH name is followed by its star rating, lines of
code, and the number of occurrences. As with conversion, the projects that
define the most implicits are not necessarily the ones with most calls.

  \begin{table}[!h]%
    \caption{Top implicit parameters}\label{ip}%
    \centering \footnotesize%
\begin{tabular}{lrlr}
  \hline
Project & Declarations & Project  & Callsites \\ 
  \hline
\href{https://github.com/lampepfl/dotty/}{lampepfl/dotty} (3K, 89K) & 4K & \href{https://github.com/scalatest/scalatest/}{scalatest/scalatest} (782, 76K) & 242K \\ 
  \href{https://github.com/scalaz/scalaz/}{scalaz/scalaz} (4K, 34K) & 4K & \href{https://github.com/apache/spark/}{apache/spark} (21K, 238K) & 59K \\ 
  \href{https://github.com/typelevel/cats/}{typelevel/cats} (3K, 24K) & 3K & \href{https://github.com/typelevel/cats/}{typelevel/cats} (3K, 24K) & 53K \\ 
  \href{https://github.com/robertofischer/hackerrank/}{robertofischer/hackerrank} (0, 50K) & 2K & \href{https://github.com/CommBank/grimlock/}{CommBank/grimlock} (29, 22K) & 52K \\ 
  \href{https://github.com/scalatest/scalatest/}{scalatest/scalatest} (782, 76K) & 2K & \href{https://github.com/exoego/aws-sdk-scalajs-facade/}{exoego/aws-sdk-scalajs-facade} (3, 302K) & 49K \\ 
  \href{https://github.com/sirthias/parboiled2/}{sirthias/parboiled2} (604, 6K) & 1K & \href{https://github.com/akka/akka/}{akka/akka} (10K, 109K) & 43K \\ 
  \href{https://github.com/laserdisc-io/laserdisc/}{laserdisc-io/laserdisc} (23, 7K) & 1K & \href{https://github.com/monix/monix/}{monix/monix} (1K, 34K) & 40K \\ 
  \href{https://github.com/slamdata/quasar/}{slamdata/quasar} (742, 27K) & 1K & \href{https://github.com/scalaz/scalaz/}{scalaz/scalaz} (4K, 34K) & 39K \\ 
  \href{https://github.com/etorreborre/specs2/}{etorreborre/specs2} (642, 26K) & 984 & \href{https://github.com/slamdata/quasar/}{slamdata/quasar} (742, 27K) & 31K \\ 
  \href{https://github.com/EHRI/ehri-frontend/}{EHRI/ehri-frontend} (10, 68K) & 981 & \href{https://github.com/lampepfl/dotty/}{lampepfl/dotty} (3K, 89K) & 29K \\ 
   \hline
\end{tabular}
  \end{table}%

\noindent
Calls sites with implicit parameters are frequent, they account for
\TgIpCallsitesPct (\TgIpCallsitesRnd) of all \Scala call sites. As shown in
Table~\ref{p}, tests account for 58\% of these calls. Small applications have a
lower proportion, most likely because they account for relatively few lines of
code.

In terms of declarations, \TgProjectsDefiningImplicitsPct of projects
(\TgProjectsDefiningImplicitsRnd) have over \TgImplicitDeclarationsRnd implicit
parameter declarations. The remaining projects do not declare any. The majority,
\TgImplicitDeclarationsPublicPct (\TgImplicitDeclarationsPublicRnd), of
declarations are public. Over half of the declarations come from 200 projects
which often implement DSL-like APIs. This also happens internally in
applications. For example, \c{ornicar/lila}, an open source chess server, is
one of the largest and most popular apps in the corpus. It uses implicits for a
small database management DSL.

\subsection{Idioms and Patterns}

In this subsection, we look at popular implicit idioms and answer the question
how frequently are these idioms used. For each, we describe the heuristic used
to recognize the pattern and give a table with the 10 top most projects in
terms of declarations as well as in use in terms of call sites. Each of the
table has the same structure: each project’s GitHub name is followed by its
star rating, lines of code, and the number of occurrences for declarations and
call sites.

Table~\ref{iap} gives a summary of the declaration and uses of the various
idioms and patterns split by our code categories; $X\ (Y\%\ Z\%)$ are such
that $X$ is the number of occurrences, $Y\%$ is the ratio of $X$ over all
categories and $Z$ is a ratio of projects in the given category.

\begin{table}[!h]
  \caption{Idioms and  patterns} \label{iap}
  \centering \small
  \begin{minipage}{\linewidth}
\begin{tabularx}{\textwidth}{Xrrrr}
  \hline
Pattern & Small Apps & Large Apps & Libraries & Tests \\ 
  \hline
Late Trait Implementation & 278 (08\% 04\%) & 968 (28\% 15\%) & 2.1K (59\% 14\%) & 177 (05\% 01\%) \\ 
  Extension Methods & 1.7K (09\% 17\%) & 5.1K (28\% 48\%) & 10.5K (57\% 45\%) & 1.2K (06\% 08\%) \\ 
  Type Classess & 4.3K (05\% 19\%) & 17.2K (21\% 49\%) & 54.2K (67\% 53\%) & 5.8K (07\% 15\%) \\ 
  Extension Syntax Methos & 1.3K (06\% 09\%) & 4.3K (20\% 28\%) & 13.9K (66\% 31\%) & 1.6K (08\% 06\%) \\ 
  Type Proofs & 110 (06\% 01\%) & 320 (18\% 05\%) & 1.3K (73\% 06\%) & 39 (02\% 00\%) \\ 
  Context & 5K (06\% 25\%) & 34.9K (41\% 62\%) & 39.2K (46\% 50\%) & 5.7K (07\% 14\%) \\ 
  Unrelated Conversions & 672 (02\% 07\%) & 2.3K (06\% 26\%) & 38.1K (92\% 20\%) & 441 (01\% 03\%) \\ 
  Bidirectional Conversion & 197 (17\% 01\%) & 321 (28\% 06\%) & 556 (49\% 03\%) & 61 (05\% 00\%) \\ 
   \hline
\end{tabularx}
  \\[1mm]
    \centering \textsf{(a) Declarations}
  \end{minipage}  \\[3mm]
  \begin{minipage}{\linewidth}
\begin{tabularx}{\textwidth}{Xrrrr}
  \hline
Pattern & Small Apps & Large Apps & Libraries & Tests \\ 
  \hline
Late Trait Implementation & 21.4K (07\% 54\%) & 67.3K (22\% 84\%) & 97.8K (31\% 54\%) & 125.4K (40\% 47\%) \\ 
  Extension Methods & 40.9K (03\% 68\%) & 207.7K (13\% 95\%) & 250.7K (15\% 82\%) & 1.1M (69\% 90\%) \\ 
  Type Classess & 99.4K (05\% 86\%) & 502.2K (23\% 99\%) & 544K (25\% 92\%) & 1.1M (48\% 88\%) \\ 
  Extension Syntax Methos & 42.7K (03\% 55\%) & 213.5K (16\% 89\%) & 227.5K (17\% 61\%) & 881K (65\% 75\%) \\ 
  Type Proofs & 1.7K (03\% 19\%) & 10.6K (19\% 61\%) & 14.9K (27\% 44\%) & 28.8K (51\% 19\%) \\ 
  Context & 35.9K (02\% 60\%) & 239.2K (14\% 87\%) & 154.6K (09\% 61\%) & 1.3M (75\% 84\%) \\ 
  Unrelated Conversions & 29.7K (07\% 72\%) & 107.4K (25\% 96\%) & 112.9K (26\% 78\%) & 178.1K (42\% 57\%) \\ 
  Bidirectional Conversion & 1.9K (06\% 13\%) & 7.9K (25\% 42\%) & 8.8K (28\% 26\%) & 13.2K (41\% 13\%) \\ 
   \hline
\end{tabularx}
    \\[1mm]
    \centering \textsf{(b) Call sites}
  \end{minipage}  \\[3mm]
\end{table}

\subsubsection{Late Trait Implementation}

Late traits are recognized by looking for \c{implicit def T=>R} where \c{R}
is a \Scala trait or \Java interface.  Technically, the same effect can be
achieved with an \c{implicit class} extending a trait, but in all cases the
implicit class adds additional methods, and thus is disqualified.  As
Table~\ref{iap} shows there are only a few declarations of this pattern,
mostly in libraries.  Table~\ref{lt} gives the top 10 projects using late
traits.

  \begin{table}[!h]%
    \caption{Top late traits}\label{lt}%
    \centering \footnotesize%
\begin{tabular}{lrlr}
  \hline
Project & Declarations & Project  & Callsites \\ 
  \hline
\href{https://github.com/lift/framework/}{lift/framework} (1K, 42K) & 152 & \href{https://github.com/exoego/aws-sdk-scalajs-facade/}{exoego/aws-sdk-scalajs-facade} (3, 302K) & 49K \\ 
  \href{https://github.com/lampepfl/dotty/}{lampepfl/dotty} (3K, 89K) & 106 & \href{https://github.com/scalatest/scalatest/}{scalatest/scalatest} (782, 76K) & 9K \\ 
  \href{https://github.com/etorreborre/specs2/}{etorreborre/specs2} (642, 26K) & 94 & \href{https://github.com/akka/akka/}{akka/akka} (10K, 109K) & 6K \\ 
  \href{https://github.com/scala/scala/}{scala/scala} (11K, 139K) & 82 & \href{https://github.com/CommBank/grimlock/}{CommBank/grimlock} (29, 22K) & 4K \\ 
  \href{https://github.com/CommBank/grimlock/}{CommBank/grimlock} (29, 22K) & 81 & \href{https://github.com/hmrc/tai/}{hmrc/tai} (1, 13K) & 3K \\ 
  \href{https://github.com/scalatest/scalatest/}{scalatest/scalatest} (782, 76K) & 74 & \href{https://github.com/broadinstitute/cromwell/}{broadinstitute/cromwell} (384, 65K) & 3K \\ 
  \href{https://github.com/l-space/l-space/}{l-space/l-space} (3, 17K) & 68 & \href{https://github.com/maif/izanami/}{maif/izanami} (91, 19K) & 2K \\ 
  \href{https://github.com/anskarl/auxlib/}{anskarl/auxlib} (1, 1K) & 63 & \href{https://github.com/etorreborre/specs2/}{etorreborre/specs2} (642, 26K) & 2K \\ 
  \href{https://github.com/anskarl/LoMRF/}{anskarl/LoMRF} (58, 13K) & 63 & \href{https://github.com/mattpap/mathematica-parser/}{mattpap/mathematica-parser} (24, 476) & 2K \\ 
  \href{https://github.com/squeryl/squeryl/}{squeryl/squeryl} (521, 9K) & 49 & \href{https://github.com/playframework/play-json/}{playframework/play-json} (193, 5K) & 2K \\ 
   \hline
\end{tabular}
  \end{table}%

\noindent
Most conversions, \TgLtGoodPct, are used between types defined in the same
project. Conditional implementation account for \TgLtConditionalPct of this
pattern. \TgLtJavaPct convert \Java types (from \TgLtJavaLibraries different
libraries). Focusing on the JDK, \TgLtJavaIo conversions are related to I/O,
\TgLtJavaLang are from \Java primitives and \TgLtJavaDate involve time and
date types. There are \TgLtFromPrimitive conversions from \Scala primitives
with \c{String} (\TgLtFromString) and \c{Int} (\TgLtFromInt) being the most
often converted from.

\subsubsection{Extension Methods}

In general extension methods can be defined using both \c{implicit class}
and \c{implicit def}. While the former is preferred, the latter is still
being used. Since an \c{implicit def} can be also used for late trait
implementation or to simply relating two types, we only consider \c{implicit
  def} with a return type that is neither a \Scala trait nor a \Java
interface and that is defined in the same file as the conversion target
because extension methods are usually collocated in either the same
compilation unit or in the source file. We found \TgEmClassOverDef implicit
classes, \TgEmClassOverDefPct of all extension methods.  Table~\ref{iap}
shows that extension methods are widely used, they are defined across the
corpus and in particular in large applications and libraries. Their use is
widespread as well. The top 10 projects using extension methods appear in
Table~\ref{em}.

  \begin{table}[!h]%
    \caption{Top extension methods}\label{em}%
    \centering \footnotesize%
\begin{tabular}{lrlr}
  \hline
Project & Declarations & Project  & Callsites \\ 
  \hline
\href{https://github.com/pbaun/rere/}{pbaun/rere} (4, 14K) & 428 & \href{https://github.com/scalatest/scalatest/}{scalatest/scalatest} (782, 76K) & 87K \\ 
  \href{https://github.com/etorreborre/specs2/}{etorreborre/specs2} (642, 26K) & 295 & \href{https://github.com/exoego/aws-sdk-scalajs-facade/}{exoego/aws-sdk-scalajs-facade} (3, 302K) & 46K \\ 
  \href{https://github.com/scalaz/scalaz/}{scalaz/scalaz} (4K, 34K) & 281 & \href{https://github.com/apache/spark/}{apache/spark} (21K, 238K) & 24K \\ 
  \href{https://github.com/scalan/special/}{scalan/special} (2, 33K) & 248 & \href{https://github.com/akka/akka/}{akka/akka} (10K, 109K) & 22K \\ 
  \href{https://github.com/lampepfl/dotty/}{lampepfl/dotty} (3K, 89K) & 214 & \href{https://github.com/hmrc/tai-frontend/}{hmrc/tai-frontend} (0, 31K) & 14K \\ 
  \href{https://github.com/ritschwumm/scutil/}{ritschwumm/scutil} (6, 12K) & 214 & \href{https://github.com/getquill/quill/}{getquill/quill} (1K, 11K) & 13K \\ 
  \href{https://github.com/typelevel/cats/}{typelevel/cats} (3K, 24K) & 171 & \href{https://github.com/hmrc/tai/}{hmrc/tai} (1, 13K) & 13K \\ 
  \href{https://github.com/lift/framework/}{lift/framework} (1K, 42K) & 168 & \href{https://github.com/monix/monix/}{monix/monix} (1K, 34K) & 12K \\ 
  \href{https://github.com/broadinstitute/cromwell/}{broadinstitute/cromwell} (384, 65K) & 166 & \href{https://github.com/broadinstitute/cromwell/}{broadinstitute/cromwell} (384, 65K) & 10K \\ 
  \href{https://github.com/monsantoco/aws2scala/}{monsantoco/aws2scala} (19, 10K) & 134 & \href{https://github.com/hmrc/iht-frontend/}{hmrc/iht-frontend} (1, 49K) & 10K \\ 
   \hline
\end{tabular}
  \end{table}%

\noindent
There are \TgEmConditionalRnd conditional extensions (\TgEmConditionalPct).
From these, \TgEmConditionalTcRnd are related to type classes and
\TgEmConditionalCtx to contexts. \TgEmJavaRnd instances extends \Java types
(\TgEmJavaPct) across \TgEmJavaLibraries libraries. Similarly to late traits,
the \Java I/O (\TgEmJavaIo), date and time (\TgEmJavaDate) and \Java primitives
(\TgEmJavaLang) are the most often extended. Extension methods are also used to
extends \Scala primitives (\TgEmFromPrimitiveRnd), again \c{String} and \c{Int}
being the most popular (\TgEmFromString and \TgEmFromInt respectively). This is
understandable as these are the basic types for building embedded DSL.

\subsubsection{Type Classes}\label{sec:analysis-type-classes}

We recognize type classes from their instances that are injected by a
compiler as implicit arguments. What differentiate them from an implicit
argument is the presence of type arguments linked to type parameters
available in the parent context. This is what distinguishes a type class and
a context.  For example, the following do not match:
\begin{lstlisting}
  def f(x: Int)(implicit y: A[Int])     def f[T](x: T)(implicit y: T)
\end{lstlisting}
while the following do:
\begin{lstlisting}
  def f[T](x: T)(implicit y: A[T])      implicit class C[T](x: T)(implicit y: A[T])
\end{lstlisting}
We match implicit parameters with at least one type argument referencing a
type parameter. Table~\ref{iap} shows that type classes are the most widely
declared pattern. Both libraries and large application use it frequently.
They are also the most frequent call sites. The top 10 projects using type
classes are in Table~\ref{tc}.

  \begin{table}[!h]%
    \caption{Top type classes}\label{tc}%
    \centering \footnotesize%
\begin{tabular}{lrlr}
  \hline
Project & Declarations & Project  & Callsites \\ 
  \hline
\href{https://github.com/scalaz/scalaz/}{scalaz/scalaz} (4K, 34K) & 4K & \href{https://github.com/scalatest/scalatest/}{scalatest/scalatest} (782, 76K) & 96K \\ 
  \href{https://github.com/typelevel/cats/}{typelevel/cats} (3K, 24K) & 3K & \href{https://github.com/exoego/aws-sdk-scalajs-facade/}{exoego/aws-sdk-scalajs-facade} (3, 302K) & 49K \\ 
  \href{https://github.com/robertofischer/hackerrank/}{robertofischer/hackerrank} (0, 50K) & 2K & \href{https://github.com/typelevel/cats/}{typelevel/cats} (3K, 24K) & 48K \\ 
  \href{https://github.com/sirthias/parboiled2/}{sirthias/parboiled2} (604, 6K) & 1K & \href{https://github.com/apache/spark/}{apache/spark} (21K, 238K) & 46K \\ 
  \href{https://github.com/slamdata/quasar/}{slamdata/quasar} (742, 27K) & 1K & \href{https://github.com/CommBank/grimlock/}{CommBank/grimlock} (29, 22K) & 43K \\ 
  \href{https://github.com/laserdisc-io/laserdisc/}{laserdisc-io/laserdisc} (23, 7K) & 1K & \href{https://github.com/scalaz/scalaz/}{scalaz/scalaz} (4K, 34K) & 38K \\ 
  \href{https://github.com/scalatest/scalatest/}{scalatest/scalatest} (782, 76K) & 947 & \href{https://github.com/slamdata/quasar/}{slamdata/quasar} (742, 27K) & 30K \\ 
  \href{https://github.com/twitter/algebird/}{twitter/algebird} (2K, 24K) & 899 & \href{https://github.com/laserdisc-io/laserdisc/}{laserdisc-io/laserdisc} (23, 7K) & 18K \\ 
  \href{https://github.com/scalanlp/breeze/}{scalanlp/breeze} (3K, 36K) & 887 & \href{https://github.com/scalaprops/scalaprops/}{scalaprops/scalaprops} (226, 6K) & 17K \\ 
  \href{https://github.com/nrinaudo/kantan.csv/}{nrinaudo/kantan.csv} (244, 5K) & 832 & \href{https://github.com/nrinaudo/kantan.csv/}{nrinaudo/kantan.csv} (244, 5K) & 16K \\ 
   \hline
\end{tabular}
  \end{table}%

\noindent
Type classes are involved in 30\% of the implicit calls which use over 11K
type classes.  Type classes are dominated by the standard library (42\%). As
expected, most come from the collection framework, \c{scala.Predef} and the
\c{math} library. Next are testing libraries (15\%) followed by the some of
the most popular frameworks and libraries including \c{Typelevel cats} and
\Scalaz that provide basic abstractions for functional programming,
including a number of common type classes. These two libraries are used by
almost 40\% in the corpus.

\subsubsection{Extension Syntax Methods}
From extension methods we select instances that define implicit parameters that
match out type class definition from Section~\ref{sec:analysis-type-classes}.
Summary is in Table~\ref{esm}.

  \begin{table}[!h]%
    \caption{Top extension syntax methods}\label{esm}%
    \centering \footnotesize%
\begin{tabular}{lrlr}
  \hline
Project & Declarations & Project  & Callsites \\ 
  \hline
\href{https://github.com/pbaun/rere/}{pbaun/rere} (4, 14K) & 428 & \href{https://github.com/scalatest/scalatest/}{scalatest/scalatest} (782, 76K) & 87K \\ 
  \href{https://github.com/etorreborre/specs2/}{etorreborre/specs2} (642, 26K) & 295 & \href{https://github.com/exoego/aws-sdk-scalajs-facade/}{exoego/aws-sdk-scalajs-facade} (3, 302K) & 46K \\ 
  \href{https://github.com/scalaz/scalaz/}{scalaz/scalaz} (4K, 34K) & 281 & \href{https://github.com/apache/spark/}{apache/spark} (21K, 238K) & 24K \\ 
  \href{https://github.com/scalan/special/}{scalan/special} (2, 33K) & 248 & \href{https://github.com/akka/akka/}{akka/akka} (10K, 109K) & 22K \\ 
  \href{https://github.com/lampepfl/dotty/}{lampepfl/dotty} (3K, 89K) & 214 & \href{https://github.com/hmrc/tai-frontend/}{hmrc/tai-frontend} (0, 31K) & 14K \\ 
  \href{https://github.com/ritschwumm/scutil/}{ritschwumm/scutil} (6, 12K) & 214 & \href{https://github.com/getquill/quill/}{getquill/quill} (1K, 11K) & 13K \\ 
  \href{https://github.com/typelevel/cats/}{typelevel/cats} (3K, 24K) & 171 & \href{https://github.com/hmrc/tai/}{hmrc/tai} (1, 13K) & 13K \\ 
  \href{https://github.com/lift/framework/}{lift/framework} (1K, 42K) & 168 & \href{https://github.com/monix/monix/}{monix/monix} (1K, 34K) & 12K \\ 
  \href{https://github.com/broadinstitute/cromwell/}{broadinstitute/cromwell} (384, 65K) & 166 & \href{https://github.com/broadinstitute/cromwell/}{broadinstitute/cromwell} (384, 65K) & 10K \\ 
  \href{https://github.com/monsantoco/aws2scala/}{monsantoco/aws2scala} (19, 10K) & 134 & \href{https://github.com/hmrc/iht-frontend/}{hmrc/iht-frontend} (1, 49K) & 10K \\ 
   \hline
\end{tabular}
  \end{table}%

We found 18.6K of syntax methods instances in 2.5K projects. Most of them are
defining operations of generic algebraic data types.

\subsubsection{Type Proofs}

We recognize this pattern by select \c{implicit def} that take generalized
type constraints, such as equality (\c{=:=}), subset (\c{<:<}) and
application (\c{=>}) as implicit type parameters. Summary
is in Table~\ref{tp}.

  \begin{table}[!h]%
    \caption{Top type proofs}\label{tp}%
    \centering \footnotesize%
\begin{tabular}{lrlr}
  \hline
Project & Declarations & Project  & Callsites \\ 
  \hline
\href{https://github.com/scalatest/scalatest/}{scalatest/scalatest} (782, 76K) & 167 & \href{https://github.com/CommBank/grimlock/}{CommBank/grimlock} (29, 22K) & 5K \\ 
  \href{https://github.com/scalikejdbc/scalikejdbc/}{scalikejdbc/scalikejdbc} (982, 13K) & 91 & \href{https://github.com/akka/akka/}{akka/akka} (10K, 109K) & 3K \\ 
  \href{https://github.com/scalanlp/breeze/}{scalanlp/breeze} (3K, 36K) & 67 & \href{https://github.com/typelevel/cats/}{typelevel/cats} (3K, 24K) & 2K \\ 
  \href{https://github.com/mpollmeier/gremlin-scala/}{mpollmeier/gremlin-scala} (412, 2K) & 54 & \href{https://github.com/outworkers/phantom/}{outworkers/phantom} (1K, 12K) & 2K \\ 
  \href{https://github.com/playframework/play-json/}{playframework/play-json} (193, 5K) & 45 & \href{https://github.com/scalatest/scalatest/}{scalatest/scalatest} (782, 76K) & 2K \\ 
  \href{https://github.com/xuwei-k/applybuilder/}{xuwei-k/applybuilder} (7, 767) & 42 & \href{https://github.com/sisioh/aws4s/}{sisioh/aws4s} (7, 15K) & 2K \\ 
  \href{https://github.com/japgolly/test-state/}{japgolly/test-state} (108, 6K) & 39 & \href{https://github.com/laserdisc-io/laserdisc/}{laserdisc-io/laserdisc} (23, 7K) & 1K \\ 
  \href{https://github.com/NICTA/scoobi/}{NICTA/scoobi} (487, 13K) & 34 & \href{https://github.com/apache/spark/}{apache/spark} (21K, 238K) & 623 \\ 
  \href{https://github.com/scoundrel-tech/scoundrel/}{scoundrel-tech/scoundrel} (0, 10K) & 34 & \href{https://github.com/tixxit/framian/}{tixxit/framian} (118, 7K) & 546 \\ 
  \href{https://github.com/twitter/scalding/}{twitter/scalding} (3K, 29K) & 34 & \href{https://github.com/scoundrel-tech/scoundrel/}{scoundrel-tech/scoundrel} (0, 10K) & 541 \\ 
   \hline
\end{tabular}
  \end{table}%

This revealed a very few projects (270) besides \Scala itself and related
projects (the new \Scala 3 compiler). They define 1.6K methods taking type
proofs as implicit parameters. Most of them are small applications which seem
to be projects experimenting with type level programming.
There are however interesting use cases. Manually inspecting the bigger
projects we found common use cases, both are related to enforcing certain API
restrictions at compile time. In one case (\ScalaJSReact{}---another project
bringing React application development into \Scala), it is used to ensure that
a given method is called only once. Another instance (\Finagle, an RPC system)
creates a type-safe builder pattern that throws a compile-time error in the
case the constructed object is missing required field. In both cases authors
used \c{@implicitNotFound} annotation to provide customized error message.

\subsubsection{Context}\label{sec:analysis-context}

Whether or not an implicit argument is an instance of the context pattern is
hard to quantify, since it depends on intent. We recognize them by selecting
implicit call sites that are neither labeled as a type class application nor as
a type proof. Summary is in Table~\ref{ctx}.

  \begin{table}[!h]%
    \caption{Top context}\label{ctx}%
    \centering \footnotesize%
\begin{tabular}{lrlr}
  \hline
Project & Declarations & Project  & Callsites \\ 
  \hline
\href{https://github.com/lampepfl/dotty/}{lampepfl/dotty} (3K, 89K) & 4K & \href{https://github.com/scalatest/scalatest/}{scalatest/scalatest} (782, 76K) & 201K \\ 
  \href{https://github.com/scalatest/scalatest/}{scalatest/scalatest} (782, 76K) & 1K & \href{https://github.com/apache/spark/}{apache/spark} (21K, 238K) & 38K \\ 
  \href{https://github.com/sirthias/parboiled2/}{sirthias/parboiled2} (604, 6K) & 1K & \href{https://github.com/akka/akka/}{akka/akka} (10K, 109K) & 28K \\ 
  \href{https://github.com/EHRI/ehri-frontend/}{EHRI/ehri-frontend} (10, 68K) & 779 & \href{https://github.com/monix/monix/}{monix/monix} (1K, 34K) & 27K \\ 
  \href{https://github.com/ornicar/lila/}{ornicar/lila} (5K, 175K) & 774 & \href{https://github.com/lampepfl/dotty/}{lampepfl/dotty} (3K, 89K) & 26K \\ 
  \href{https://github.com/ponkotuy/MyFleetGirls/}{ponkotuy/MyFleetGirls} (86, 26K) & 717 & \href{https://github.com/CommBank/grimlock/}{CommBank/grimlock} (29, 22K) & 18K \\ 
  \href{https://github.com/Sciss/SoundProcesses/}{Sciss/SoundProcesses} (23, 13K) & 715 & \href{https://github.com/hmrc/iht-frontend/}{hmrc/iht-frontend} (1, 49K) & 18K \\ 
  \href{https://github.com/sciss/fscape-next/}{sciss/fscape-next} (6, 27K) & 696 & \href{https://github.com/hmrc/tai-frontend/}{hmrc/tai-frontend} (0, 31K) & 17K \\ 
  \href{https://github.com/ruimo/store/}{ruimo/store} (5, 38K) & 688 & \href{https://github.com/gapt/gapt/}{gapt/gapt} (48, 68K) & 16K \\ 
  \href{https://github.com/sciss/patterns/}{sciss/patterns} (1, 8K) & 620 & \href{https://github.com/twitter/finagle/}{twitter/finagle} (7K, 64K) & 13K \\ 
   \hline
\end{tabular}
  \end{table}%

As expected, contexts are used heavily in projects such as \Scala compiler
(\Dotty is the new Scala compiler), \Spark or \Akka, \Ie projects that are
centered around some main context which is being passed around in number of
methods. \Java types are also used as context parameters. Together
\TgCtxJdkTypes types from JDK are used in \TgCtxJdkRnd methods across
\TgCtxJdkProjects projects. The top used one is \c{SQLConnection} followed by
interfaces from \c{java.io}. Scala primitive types are used in \TgCtxPrimitive
methods in \TgCtxPrimitiveProjects. Function types are also used as contexts
(\TgCtxFunction methods in \TgCtxFunctionProjects projects), providing a convenient way
to define application counters, implicit filters and other default data
processors.

\subsubsection{Anti-pattern: Conversions}

Unrelated conversions are public, top-level definitions defined outside of
either from or to compilation units. We recognize them by selecting implicit
conversions that are not block-local, or private, or protected and are not
defined in the same compilation unit as the source or the target type. Summary
is in Table~\ref{uic}.

  \begin{table}[!h]%
    \caption{Top unrelated conversions}\label{uic}%
    \centering \footnotesize%
\begin{tabular}{lrlr}
  \hline
Project & Declarations & Project  & Callsites \\ 
  \hline
\href{https://github.com/shadaj/slinky/}{shadaj/slinky} (265, 46K) & 34K & \href{https://github.com/apache/spark/}{apache/spark} (21K, 238K) & 21K \\ 
  \href{https://github.com/sisioh/aws4s/}{sisioh/aws4s} (7, 15K) & 402 & \href{https://github.com/gapt/gapt/}{gapt/gapt} (48, 68K) & 11K \\ 
  \href{https://github.com/CommBank/grimlock/}{CommBank/grimlock} (29, 22K) & 299 & \href{https://github.com/scalatest/scalatest/}{scalatest/scalatest} (782, 76K) & 8K \\ 
  \href{https://github.com/scala/scala/}{scala/scala} (11K, 139K) & 166 & \href{https://github.com/akka/akka/}{akka/akka} (10K, 109K) & 7K \\ 
  \href{https://github.com/etorreborre/specs2/}{etorreborre/specs2} (642, 26K) & 130 & \href{https://github.com/CommBank/grimlock/}{CommBank/grimlock} (29, 22K) & 6K \\ 
  \href{https://github.com/squeryl/squeryl/}{squeryl/squeryl} (521, 9K) & 128 & \href{https://github.com/ornicar/lila/}{ornicar/lila} (5K, 175K) & 6K \\ 
  \href{https://github.com/scoundrel-tech/scoundrel/}{scoundrel-tech/scoundrel} (0, 10K) & 113 & \href{https://github.com/ilya-klyuchnikov/tapl-scala/}{ilya-klyuchnikov/tapl-scala} (126, 13K) & 3K \\ 
  \href{https://github.com/scala/scala-java8-compat/}{scala/scala-java8-compat} (353, 4K) & 112 & \href{https://github.com/scoundrel-tech/scoundrel/}{scoundrel-tech/scoundrel} (0, 10K) & 3K \\ 
  \href{https://github.com/typelevel/cats/}{typelevel/cats} (3K, 24K) & 110 & \href{https://github.com/broadinstitute/cromwell/}{broadinstitute/cromwell} (384, 65K) & 2K \\ 
  \href{https://github.com/lift/framework/}{lift/framework} (1K, 42K) & 101 & \href{https://github.com/mattpap/mathematica-parser/}{mattpap/mathematica-parser} (24, 476) & 2K \\ 
   \hline
\end{tabular}
  \end{table}%

There are \TgUicDefinedRnd of unrelated conversions spanning across
\TgUicDefiningProjectsRnd projects (\TgUicDefiningProjectsPct). Most of them
(\TgUicDefinedInSlinkyRnd) belong to the already mentioned \Slinky projects
bringing React apps development to Scala. They are used in
\TgUicUsingProjectsRnd (\TgUicUsingProjectsPct) projects. If we change the query
to only regard the same artifact then it drops too \TgUicDefinedInLibsRnd
conversions in \TgUicDefinedInLibsProjects projects. There are some indication
that unrelated conversions might be deprecated in the upcoming revision of the
\Scala language\footnote{\Cf
  \url{https://github.com/lampepfl/dotty/pull/2060}}. The numbers here show,
that these conversions are being defined, but they are usually in the scope of
the same library.
From the unrelated conversions, \TgUicPrimitiveRnd from
\TgUicPrimitiveProjects projects involves \Scala primitive types. They are
present in all categories, but majority comes from libraries where they are
used as building blocks for DSLs. Only a very few (\TgUicBothPrimitive in
\TgUicBothPrimitiveProjects projects) convert just between primitive types.

For the conversions that go both ways, we consider all pairs of such conversions
that are defined in the same artifact and thus could be easily imported in the
same scope.
%

We have identified \TgBicDefinedRnd such conversions defined in
\TgBicDefiningProjects (\TgBicDefiningProjectsPct) projects and used across
\TgBicUsingProjectsRnd (\TgBicUsingProjectsPct) projects. As expected, this has
matched all the Scala-Java collection conversions defined in the
\c{scala.collection} package. They are used in \TgBicProjectsUsingWrapAs
(\TgBicProjectsUsingWrapAsPct) projects. This is significantly less than the
recommended alternative using explicit \c{asJava} or \c{asScala} decorators
that are being used by \TgBicProjectsUsingDecorateAsPct of projects.
\TgBicProjectsUsingBoth projects mix both approaches.

From a manual inspection of some of the other popular bi-directional
conversion, we find that it is used in libraries that provide both Java and
Scala API (\Eg \Spark or \Akka) allowing one to freely mix Java and Scala
version of the classes. Some libraries use them to provide easier syntax for
its domain objects (\Eg using a tuple to represent a cell coordinate), or
lifting types from/to \c{scala.Option}. Another distinct category are
conversions between many of the different date and time representations in both
Java and Scala. We found only \TgBicFromOrToPrimitives bi-directional
conversions involving primitive types, out of which \TgBicFromAndToPrimitives
are only between primitives.

\subsection{Complexity}

\begin{figure}[tb]\begin{minipage}[t]{.37\linewidth}
  \includegraphics[width=\linewidth]{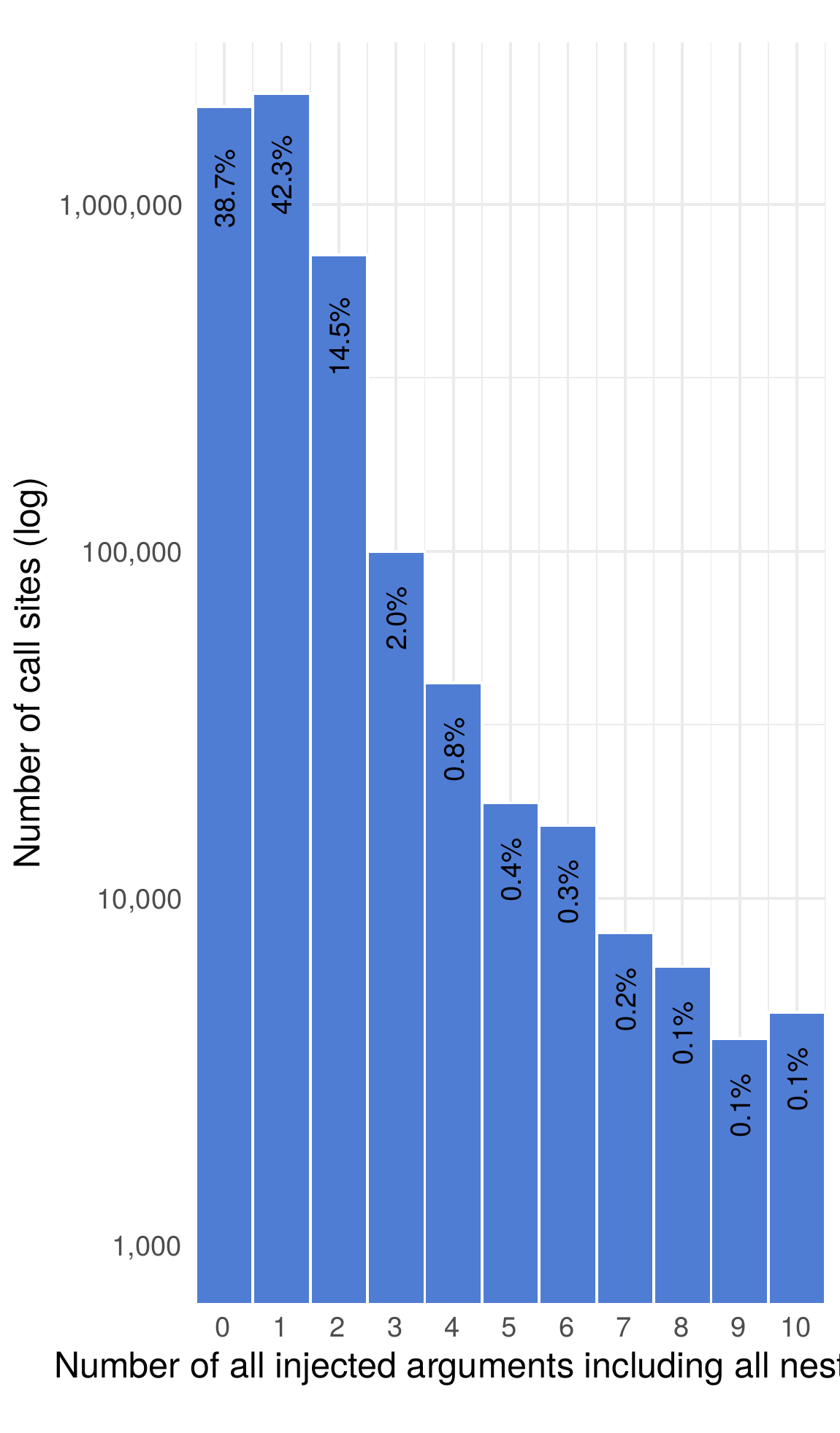}
  \caption{Injected arguments}\label{fig:hist-nested-arguments}
\end{minipage}\hfill\begin{minipage}[t]{.37\linewidth}
  \includegraphics[width=\linewidth]{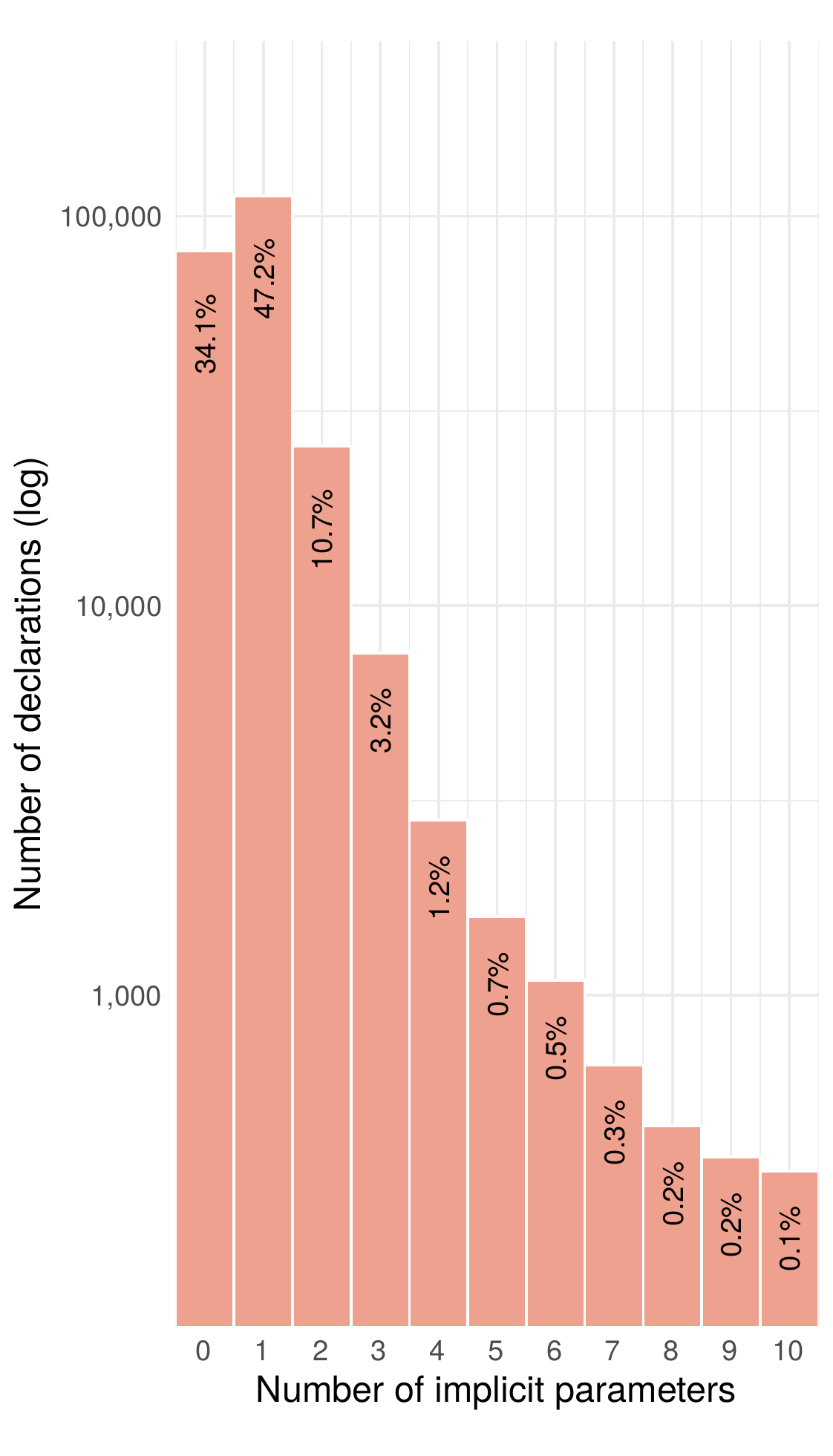}
  \caption{Implicit parameters}\label{fig:hist-impl-parameters}
\end{minipage}\end{figure}

One question we wanted to address was the amount of work performed by the
\Scala compiler. This is motivated by the need for the programmer to reverse
engineer the compiler's work to understand how to fix their code when an
error is related to implicits. In terms of code size, if one were to sum up
the length of the symbols inserted by the compiler at the various call sites
that use implicit arguments, this would amount to
\TgLengthOfInjectedDeclarationsRnd characters or about
\TgLengthOfInjectedDeclarations{}x the size of the entire \Scala project.
Figure~\ref{fig:hist-nested-arguments} shows the distribution of injected
implicit arguments into methods. We limit the graph to 10 injected
arguments, but in practice there is a long tail.  The measurements are
obtained by inspecting each call site where implicit resolution is involved
and counting arguments injected directly to the target function as well as
arguments injected to nested calls needed for the implicit derivation. While
the distribution has a long tail, going all the way to \TgCsNestingMax, the
median is \TgCsNestingMedian. At the extreme, the {\tt \TgCsNestingMaxProject}
project is exploring type-level programming and has one call site that
includes \TgCsNestingMax nested implicit calls and value
injection. Expressed in length of the injected code, that call site has the
compiler inject \TgCsNestingMaxCharLengthRnd characters.
Figure~\ref{fig:hist-impl-parameters} shows the distribution of the number
of implicit parameter declarations. The data suggests that programmers are
likely to encounter functions with one or two implicits rather frequently.
And they are likely to deal with functions with four or more implicits
several times per project.

To help navigate this complexity, the Scala plugin for Intellij IDEA has a
feature that can show implicit hints, including implicit resolution in the code
editor. This effectively reveals the injected code making it an indispensable
tool for debugging. However, turning the implicit hints on severely hinders the
editor performance creating a significant lag when working with implicits-heavy
files. The second problem with this is that the Intellij \Scala compiler is not
the same as \Scalac and implicit resolution often disagrees between compiler
implementations (\Eg Intellij does not consider implicit shadowing in lexical
context). Another way to mitigate some of the complexity related to errors
occurring during resolution is to customize the error message emitted when an
implicit type is not found. \Scala provides the \c{@implicitNotFound(message)}
annotation to this end, where \c{message} can be parameterized with the names
of type parameters that the type defines. In the corpus, we have found it
defined \TgImplicitNotFoundTypesRnd times in \TgImplicitNotFoundProjects
projects, and used in \TgImplicitNotFoundUseRnd call sites.

\subsection{Overheads}

\begin{figure}[!b]
  \includegraphics[width=\textwidth]{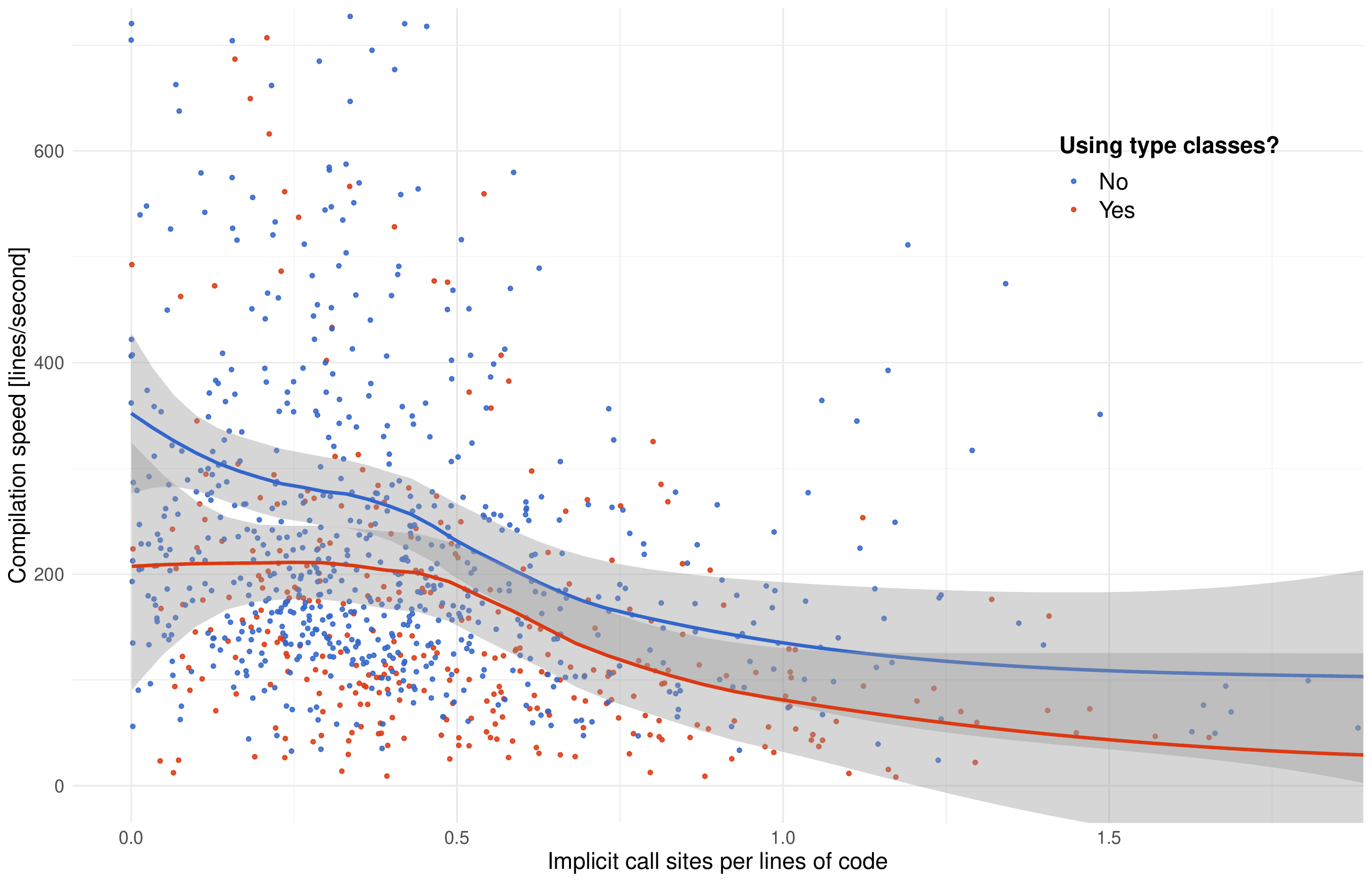}
  \caption{Compilation slowdown}\label{fig:compilation-speed}
\end{figure}

Another question we are interested to investigate is the effect of implicits
on compile time. We have demonstrated that on a synthetic example, resolution
can significantly impact type-checking performance.  There are
\CompilationTimeProjects (\CompilationTimeProjectsCode LOC) using \Scala
2.12.4+ for which we can get compile time statistics using the
\c{-Ystatistics:typer} compiler flag. Furthermore \CompilationTimeProjectsTC
projects (\CompilationTimeProjectsTCCode LOC) use the \c{shapeless} library
which is the most common approach to guide the type class
derivation~\cite{canteor18_scalac_profiling}. The result of measuring
compilation speed between these two sets of projects is shown in
Figure~\ref{fig:compilation-speed}. More precisely, the figure shows data
for projects that have more than 1,000 lines of code (for smaller projects
compilation times may be dominated by startup costs). The x-axis shows the
density of implicit call sites (their ratio per line of code, ranging
between 0 and almost 2).  The y-axis shows compilation speed measured in
lines per second. For this figure we capture the entire compilation time of
each project, including I/O. Higher is better on this graph.
Colors distinguish projects that use type classes (red) from those who do
not (blue). The lines indicate an estimate of the conditional mean function
(loess). If implicits were not influencing compilation time, one would expect
both lines to be roughly flat and at the same level. What we see instead
confirms our hypothesis, the cost of compilation increases with the density
of implicits and the use of type classes further reduce compilation speed.

Another manifestation can be found in the \ScalaTest testing framework. It
defines a \c{Prettifier} for pretty printing which looks like a perfect
candidate for a type class, yet the authors have decided to use it as a context
parameter instead. The reason given for that is performance: ``\emph{Prettifier
  is not parameterized ... because assertions would then need to look up
  Prettifiers implicitly by type. This would slow compilation.''}
In the corpus there are over \TgCtxPrettifierRnd calls to methods using the
\c{Prettifier} context. Resolving all of them implicitly using the implicit
type class derivation machinery could indeed induce a slowdown across
\TgCtxPretrifierProjectsRnd projects.

\subsection{Threats to Validity}
We report on two source of threats to validity.  One threat to
\emph{external validity} is linked to selection of code that was
analyzed. We analyzed 15\% of the \Scala code publicly-available on \GH. Our
findings only generalize to industry if the code we analyzed is
representative of industrial use of implicits. It is possible, for instance,
that some companies enforce coding guidelines that impact the usage of
implicits. We have no evidence that this is the case, but cannot rule it
out.  In terms of threats to \emph{internal validity} we consider our data
analysis pipeline. It has several sources of inaccuracies. We rely on
\ScalaMeta to gather synthetic call sites. \ScalaMeta restricts us to two
\Scala versions and it only generates metadata for about half of the
selected projects. We are also aware that for 3\% of implicit uses symbols
could not be resolved.

\section{Related Work}

The design of implicits as it appears in \Scala is but one point in a larger
space. While alternative designs are out of the scope of this work, we
mention some important related work.  \citet{oliveira10_type_classes}
established the connection between Haskell's type classes and \Scala
implicits with multiple examples.  \citet{oliveira12_implicit_calculus}
formalized the key ideas of implicits in a core calculus. \citet{rouvoet16}
expanded the Oliveira et al. work and proved soundness and partial
completeness independent of termination.  \citet{schrijvers19_cochis}
present an improved variant of the implicit calculus.  One key property of
this work is the notion of \emph{coherence} (which is attributed to
\citet{reynolds91}). Coherence requires a program to have a single meaning,
i.e. it precludes any semantic ambiguity. \Scala eschews coherence in favor
of expressivity by allowing overlapping implicits. Schrijvers et al. propose
a design that recovers coherence.

There have been efforts to study how \Scala is used by practitioners.
\citet{tasharofi13} looked at how often and why \Scala developers mix the actor
model with other models of concurrency. They analyzed only \GH 16 projects at
the compiled byte-code level with a custom tool. The choice of byte-code had
some drawbacks. For example, their analysis could not detect indirect method
invocations and thus they had to supplemented it with manual inspection. The
same corpus is used by \citet{koster15} to analyze different synchronization
mechanisms used in \Scala code. Despite using the same projects, he analyzed
80\% more lines of code as the projects were updated to their latest commit.
The increase was mostly due to \Spark that grew from 12K to 104K lines of code.
Unlike the previous study, he opted for source code analysis based on string
matching. \citet{bleser19} analyzed the tests of 164 \Scala projects (1.7M LOC)
for a diffusion of test smells. They used a similar way of assembling a corpus.
While they started with 72K projects, but only managed to compile 2.9K
projects. They discard projects with less than 1K LOC or without \ScalaTest
unit tests. For analysis, they also used semantic data from the \SDB.

\citet{pradel15} analyzed the use of implicit type conversions in
JavaScript.  They use dynamic analysis running hundreds of programs
including the common JavaScript benchmarks and popular real-world
websites. In JavaScript, implicit type conversion is basically a type
coercion. Despite that the coercion rules are well formalized, they are
fairly complex and confuse even seasoned JavaScript developers. Unlike in
\Scala that has static type system, JavaScript uses implicit type conversion
extensively (it is present in over 80\% of the studied programs), yet the
study finds that over 98\% of the conversion is what the authors consider as
harmless.

\section{Conclusions}

Implicits are a cornerstone of the \Scala programming language. There is hardly
any API without them as they enable elegant architectural design. They allow
one to remove a lot of boilerplate by leveraging the compiler's knowledge about
the code. However, they can be also easily misused and if taken too far
seriously hurt the readability of a code. Implicits are driven by type
declarations. Thus, while, implicits are \emph{used} transparently, with no
indication in the program text, their application is guided by clear and
precise rules. Our data shows that programmers have embraced them, with
\TgProjectsUsingImplicitsPct of the projects we analyzed using them, and
\TgProjectsDefiningImplicitsPct of projects defining at least one implicit
declaration. We also observed the prevalence of the idioms described, as most
projects use them in some form. For implicit conversions, our results indicate
that \TgIcAllProjectsUsingPct of projects make use of them at some point, with
the most popular conversions coming from the standard library and testing
libraries. From the idioms we presented in this paper, type classes and
extension methods are used extensively. Regarding conversions, most convert to
and from types within the scope of the project. However, there is a number of
conversions defined on unrelated types. While deprecating this form of
conversion has been discussed, doing so would break \TgUicDefiningProjectsRnd
projects (\TgUicDefiningProjectsPct) in our corpus.

\paragraph{Observations for language designers.}  We have seen many source
of complexities related to the notion of coherence.  Future designs of
implicits should strongly consider adopting some limits to expressivity in
order to improve code comprehension. A related point is to avoid relying on
names of implicits during their resolution as this leads to subtle errors.
Better tool support  and static analysis could help diagnose performance
problems and could help code comprehension, but it is crucial that
IDEs and the \Scala compiler agree on how resolution is to be performed.

\paragraph{Observations for library designers.}
Over-engineered libraries are hard to understand. It is worth considering
the costs and benefits of adding, for example, type classes to an API.
Asking questions such as ``Is retroactive extension an important use case?''
or ``How much boilerplate can actually be avoided?'' may help target the
right use-cases for implicits. Often the key design issue is whether good
defaults can be provided. When they cannot, the benefits of implicits
decrease significantly. A good library design is one that lets regular users
benefit without forcing them fully understand the cleverness that the
library designer employed. Finally, we leave designers with the following
unsolicited advice: Do not use unrelated implicits! Do not use conversions
that go both ways!  Do not use conversions that might change semantics!

\section*{Acknowledgments}

Borja Lorente Escobar implemented an early version of the pipeline presented in
this paper, we thank him for his contributions. We thank the reviewers for
constructive comments that helped us improve the presentation. We thank Ólafur
Páll Geirsson for his help with \SDB and \ScalaMeta. We thank the members of
the PRL lab in Boston and Prague for additional comments and encouragements.
This work received funding from the Office of Naval Research (ONR) award
503353, from the European Research Council under the European Union's Horizon
2020 research and innovation program (grant agreement 695412), the NSF (awards
1544542, and 1617892), and the Czech Ministry of Education, Youth and Sports
(grant agreement CZ.02.1.01\/0.0\/0.0\/15\_003\/0000421).

\newpage

\end{document}